\documentclass[reqno,12pt]{amsart}
\usepackage{fullpage}
\usepackage{amsfonts,amssymb,amsmath}
\numberwithin{equation}{section}

\newtheorem{thm}{Theorem}[section]
\newtheorem{cor}{Corollary}[section]

\newtheorem{prop}{Proposition}[section]

\numberwithin{equation}{section}
\def\Ref#1{Ref.\cite{#1}}

\def\p{\partial}

\def\mbs{\boldsymbol}
\def\tens#1{\mbs{#1}}

\def\parder#1#2{\frac{\p #1}{\p #2}}

\def\eos/{equation of state}
\def\esos/{equations of state}
\def\com/{constant of motion}
\def\csom/{constants of motion}
\def\ie/{i.e.}
\def\eg/{e.g.}

\def\ertl{{\rm E}}
\def\ross{{\rm R}}
\def\holl{{\rm H}}

\def\Rnum{\mathbb{R}}

\def\t{{\rm t}}

\def\vecder{\vec{\nabla}}
\def\grad{\text{\em grad}\;}
\def\div{\text{\em div}\;}
\def\curl{\text{\em curl}\;}

\def\Dt{\mathfrak{D}_t}
\def\dt{\frac{d}{dt}}
\def\Lu{\mathcal{L}_{\u}}
\def\lieder#1{\mathcal{L}_{#1}}
\def\hook{\rfloor}
\def\nor{\hat\nu}
\def\d{\mathbf{d}}
\def\ddot{\mkern-2mu :\mkern-2mu}

\def\e#1{\hat e_{(#1)}}
\def\voltens{{\boldsymbol{\epsilon}}}
\def\volform{\epsilon}
\def\metr{g}
\def\invmetr{\mbs{\metr}^{-1}}
\def\dx{\d\vec{x}}

\def\x{\vec{x}}
\def\u{\vec{u}}
\def\vel{{\mbs u}}
\def\vort{\vec{\omega}}
\def\symmderu{{\mbs s}}
\def\skewderu{{\mbs \omega}}

\def\inv{K}
\def\Inv{{\mathcal K}}
\def\invwedge{\Inv_0\wedge}
\def\invtimes{\rho\tens{\inv}_0\times}
\def\invdiv{(1/\rho)\div\rho}

\def\v{\vec{v}}
\def\w{\vec{w}}
\def\wvort{\vec{\varpi}}

\def\V{\mathcal{V}}
\def\S{\mathcal{S}}
\def\C{\mathcal{C}}
\def\dV{\,dV}
\def\dS{d\vec{A}}
\def\ds{d\vec{s}}
\def\dA{\,dA}
\def\dl{\,d\ell}

\begin{document}

\title{Hierarchies of new invariants and conserved integrals\\ in inviscid fluid flow}

\author{
Stephen C. Anco$^1$
\lowercase{\scshape{and}}
Gary M. Webb$^{2}$ 
\\\\\lowercase{\scshape{
${}^1$Department of Mathematics and Statistics\\
Brock University\\
St. Catharines, ON L$\scriptstyle{2}$S$\scriptstyle{3}$A$\scriptstyle{1}$, Canada}} \\\\
\lowercase{\scshape{
${}^2$Center for Space Plasma and Aeronomic Research\\
The University of Alabama in Huntsville\\
Huntsville AL $\scriptstyle{35805}$, USA}}
}

\thanks{
Email: sanco@brocku.ca, gmw0002@uah.edu\\
S.C.A.\ is supported by an NSERC research grant. 
G.M.W.\ is supported in part by NASA grant NNX15A165G\kern-1pt
}

\begin{abstract}
A vector calculus approach for the determination of 
advected invariants is presented for inviscid fluid flow in three dimensions. 
This approach describes invariants by means of Lie dragging of scalars, vectors, and skew-tensors with respect to the fluid velocity,
which has the physical meaning of characterizing tensorial quantities that are frozen into the flow. 
Several new main results are obtained. 
First, 
simple algebraic and differential operators that can be applied recursively
to derive a complete set of invariants starting from the basic known local and nonlocal invariants 
are constructed. 
Second, 
these operators are used to derive infinite hierarchies of local and nonlocal invariants 
for both adiabatic fluids and homentropic fluids that 
are either incompressible, or compressible with barotropic and non-barotropic \esos/. 
Each hierarchy is complete in the sense that no further invariants can be generated from the basic local and nonlocal invariants. 
All of the resulting new invariants are generalizations of  
Ertel's invariant, the Ertel-Rossby invariant, and Hollmann's invariant.
In particular, for incompressible fluid flow in which the density is non-constant 
across different fluid streamlines, 
a new variant of Ertel's invariant and several new variants of Hollmann's invariant
are derived, where the entropy gradient is replaced by the density gradient. 
Third, the physical meaning of these new invariants and the resulting conserved integrals 
is discussed,
and their relationship to conserved helicities and cross-helicities is described. 
\end{abstract}


\maketitle
\allowdisplaybreaks[3]

\section{Introduction}

Vorticity invariants and conserved helicity integrals 
have long been recognized to be important 
\cite{Bat-book,MajBer-book,WuZhou-book,ArnKhe-book} 
in the study of inviscid fluid flow in three dimensions, 
especially for understanding topological aspects of vortex flows 
and for studying existence, uniqueness, and stability of initial-value flows.

For a general hydrodynamical system on any spatial domain, 
an \emph{invariant} is a material quantity 
constructed from the fluid variables (and possibly their spatial derivatives)
such that it is advected by the flow. 
Physically, this means that the quantity is frozen into the flow, 
analogously to attaching it to fluid particles transported by the flow.
Geometrically, 
an invariant has the property that its advective Lie derivative vanishes,
where this derivative is defined by \cite{TurYan,Anc13,WebDasMcKHuZan,BesFri} 
$\Dt = \p_t + \Lu$ 
in terms of the Lie derivative $\Lu$ with respect to the fluid velocity $\u$. 
The advective Lie derivative can be expressed alternatively 
as the usual material derivative $\dt = \p_t + \u\cdot\vecder$ 
plus a rotation-dilation term that takes into account the tensorial nature of the quantity 
on which it acts.
The set of all invariants of a hydrodynamical system is an intrinsic (coordinate-free) part of the fundamental structure of the system. 

A \emph{vorticity invariant} in three dimensions refers to an invariant that has 
an essential dependence on the vorticity vector of the fluid flow, 
$\vort=\vecder\times\u$, 
given by the curl of the fluid velocity $\u$.
An invariant is \emph{local} if its value at each point in the spatial domain 
is determined entirely by the values of the fluid variables 
and their spatial derivatives (up to some finite differential order) 
at that point, 
and otherwise an invariant is called \emph{nonlocal}. 

Inviscid isothermal fluid flow has only one local vorticity invariant,
which is the densitized vorticity vector $(1/\rho)\vort$,
where $\rho$ is the fluid density. 
For inviscid adiabatic fluid flow, 
the densitized vorticity vector $(1/\rho)\vort$ is no longer an invariant, 
but there is a local vorticity invariant $(1/\rho)\vort\cdot\vecder S$
known as Ertel's invariant \cite{Ert}, 
where $S$ is the fluid entropy. 
In the case of incompressible fluids, in which $\rho$ is constant, 
the factor $1/\rho$ in these two invariants can be dropped.
The vorticity vector invariant is well-known to be closely connected to the helicity conservation law
$\dt \int_{\V(t)} \u\cdot\vort\dV = \oint_{\p\V(t)} (\tfrac{1}{2}|\u|^2 - p/\rho)\vort\cdot\dS$
which holds on moving volumes $\V(t)$ 
when the fluid pressure satisfies a barotropic \eos/, $p=P(\rho)$. 
Likewise, Ertel's invariant is related to the entropy circulation-flux conservation law
$\dt \int_{\S(t)} (\u\times\vecder S)\cdot\dS = \oint_{\p\S(t)} (\tfrac{1}{2}|\u|^2 -E - p/\rho)\vecder S\cdot\dS$,
where $E$ is the internal energy of the fluid 
as determined by the standard thermodynamic relation $TdS = dE +p d(1/\rho)$
with $T$ being the fluid temperature. 
Note that the helicity integral is conserved 
when the vorticity filaments are tangential to the moving boundary surface $\p\V(t)$,
and that the entropy circulation-flux is conserved 
when the entropy gradient is tangential to the moving boundary curve $\p\S(t)$. 

Apart from these well-known local vorticity invariants, 
there exist nonlocal vorticity invariants, 
which arise from Clebsch variables related Weber transformations. 
The oldest examples are the Ertel-Rossby invariant \cite{Ros}
$(1/\rho)(\u-\vecder\phi)\cdot\vort$
for barotropic fluids, 
and Hollmann's invariant \cite{Hol},  
$(1/\rho)(\u-\vecder\phi)\cdot(\vecder S\times \vecder( (1/\rho)\vort\cdot\vecder S ))$
for adiabatic fluids, 
where $\phi$ is a Clebsch variable defined by the transport equation 
$\Dt\phi= \tfrac{1}{2}|\u|^2 - E - p/\rho$.
(Recent derivations of the Ertel-Rossby invariant can be found in \Ref{WebDasMcKHuZan,BesFri}.)

The Ertel-Rossby invariant yields a conserved helicity integral 
$\dt \int_{\V(t)} (\u-\vecder\phi)\cdot\vort\dV =0$
which measures the self-linking of the vorticity filaments defined by $\vort$.
In contrast to the familiar helicity integral, 
no boundary conditions are needed on $\V(t)$ for the integral to be conserved.
Hollmann's invariant yields a conserved cross-helicity integral 
$\dt \int_{\V(t)} (\u-\vecder\phi)\cdot(\vecder S\times\vecder( (1/\rho)\vort\cdot\vecder S ))\dV=0$
which turns out to measure the mutual linking of filaments defined by the vorticities 
$\vort$ and $\vecder\times(S\vecder( (1/\rho)\vort\cdot\vecder S ))$. 

An additional nonlocal invariant arises from the introduction of a Clebsch variable $\psi$ 
defined by the transport equation $\Dt\psi= T$
given in terms of the fluid temperature. 
This leads to a nonlocal vorticity vector invariant \cite{Mob,WebDasMcKHuZan} 
$(1/\rho)\wvort$ in terms of
$\wvort=\vecder\times(\u -\psi\vecder S)$, 
which generalizes the local vorticity vector invariant $(1/\rho)\vort$ 
to inviscid adiabatic (non-barotropic) fluids.
As an important consequence, 
there is a generalization of the Ertel-Rossby helicity invariant, 
given by \cite{WebDasMcKHuZan} 
$(\u-\vecder\phi-\psi\vecder S)\cdot\wvort$, 
for adiabatic non-barotropic fluid flow.
The corresponding helicity integral
$\dt \int_{\V(t)} (\u-\vecder\phi-\psi\vecder S)\cdot\wvort\dV =0$
is conserved without boundary conditions on $\V(t)$.

Helicity measures the knotting of vortex tubes in barotropic fluid flow \cite{Mof,BerFie,MofRic}. 
The generalized helicity in adiabatic non-barotropic fluid flow 
has interesting physical applications in the formation of tornadoes (see, e.g. \cite{Ped})
and in solar magnetohydrodynamics (see, e.g. \cite{WedSkuSteRouCruFedErd}). 
Both of these helicities are related to the generalized Aharonov-Bohm effect 
for fluids and MHD obtained in \Ref{Yah2013,Yah2017,Yah2018} 
(see also \Ref{WebAnc2017,Web-book}). 
More generally, 
vorticity invariants and associated conserved integrals are important in 
atmospheric and oceanic Rossby wave dynamics, 
where there is an interplay between the planetary vorticity and the local fluid vorticity (see, e.g. \cite{Ped}). 

Most strikingly, 
as indicated in \Ref{TurYan} and further developed in \Ref{WebDasMcKHuZan},
a method based on differential forms can be used to construct, in principle, 
a hierarchy of vorticity invariants involving higher-order derivatives of the fluid variables. 
Those results give rise to several interesting open questions:
\begin{itemize}
\item
Can the method be formulated directly in terms of an explicit generating set of material operators that map invariants into invariants? 
\item
Does the resulting hierarchy of invariants provide a complete set of all invariants?
\item
For which kinds of inviscid fluid flow is the hierarchy infinite? 
\item
What is the physical meaning of the high-order invariants 
and how are they connected to conserved helicity integrals and conserved cross-helicity integrals?
\item 
Do there exist any additional basic (local or nonlocal) invariants for inviscid fluid flow?
\end{itemize}

The purpose of the present work is to answer all of these questions 
and, as an important by-product, 
derive new invariants and corresponding new conserved integrals
(which were not presented in \Ref{TurYan,WebDasMcKHuZan,BesFri})
for inviscid fluid flow in three dimensions.
Several main results are obtained. 

The local vorticity vector invariant and the related Ertel-Rossby nonlocal invariant 
will be seen to be part of respective hierarchies of local and nonlocal invariants 
in inviscid homentropic compressible fluid flow with a barotropic \eos/.
For inviscid adiabatic compressible fluid flow,
Ertel's invariant will be seen to be part of the infinite hierarchy of local invariants,
while Hollmann's invariant and the generalized Ertel-Rossby helicity invariant
will be seen to belong to an infinite hierarchy of nonlocal invariants.
The latter hierarchy contains the nonlocal vorticity vector invariant
$(1/\rho)\wvort$,
which will be shown to coincide with a nonlocal 2-form invariant
derived recently in \Ref{BesFri}. 

For each of these hierarchies there are corresponding
material conserved integrals on moving volumes, surfaces, and curves. 
Specializations of both hierarchies to constant-density fluid flow 
also will be presented. 

In addition, 
a new variant of Ertel's invariant
and several new variants of Hollmann's invariant
will be derived for incompressible fluid flow in which the density is non-constant 
across different fluid streamlines. 
In such flows the density is frozen-in and has no effect on the dynamics of the fluid velocity.
The new invariants arise from replacing the entropy gradient by the density gradient,
and they belong to enlarged infinite hierarchies of local and nonlocal invariant, respectively. 

Furthermore, all of the hierarchies of invariants, 
covering adiabatic and homentropic inviscid compressible fluid flow,
as well as constant and non-constant density inviscid incompressible fluid flow,
are shown to be complete.
In particular,
the basic local and nonlocal invariants that are known for these types of fluid flow are shown to comprise all possible invariants of lowest order type,
and as consequence the generating set of material operators 
yields all invariants of higher order type.
These higher-order invariants will be seen to comprise
Ertel-type scalars, Hollmann-type cross-helicity scalars,
vorticity-type vectors, and gradient skew-tensors. 

Thus, the results in this paper will settle the long-standing open problem of finding all invariants within the framework of Clebsch variables 
for inviscid fluid flow in $\Rnum^3$. 

To make the presentation and main results accessible to the widest audience, 
all results will be stated using familiar vector calculus operations 
(and their extension to tensor calculus).
Likewise, we will consider fluid flow in $\Rnum^3$ without boundaries.
The methods and the results have a straightforward extension to fluids
in three-dimensional curved manifolds with and without boundaries. 

In Section~\ref{prelims}, 
we discuss the different types and properties of advected invariants. 
We also review the definition and properties of the advective Lie derivative $\Dt = \p_t + \Lu$ 
as well as its relation to the material derivative,
and we briefly explain the correspondence between invariants and conserved integrals 
on moving domains given by transported volumes, surfaces, and curves. 

In Section~\ref{operations},
we formulate the basic algebraic and differential operations that can be applied to a set of scalar, vector, tensor invariants to yield further invariants.
These material operations are the vector-calculus counterparts of the differential form methods 
discussed in \Ref{TurYan,WebDasMcKHuZan,BesFri}. 
We emphasize that the formulation of material operations is a non-trivial problem 
because the vector dot product and cross product are \emph{not advected} in fluids;
in particular, the dot/cross product of two vector invariants does not yield an invariant.
This problem has not been addressed systematically in the literature. 

In Sections~\ref{locinvs} and~\ref{nonlocinvs},
we first show that the few known local and nonlocal invariants for inviscid fluid flow exhaust all possible invariants to lowest order,
and then we apply the material operations to these invariants to derive 
complete hierarchies of local and nonlocal invariants 
for both adiabatic fluid flow and homentropic fluid flow. 
These invariants physically describe frozen-in quantities given by advected 
scalars, vectors, and skew-tensors. 
We identify all of the invariants that are of vorticity type, 
and we discuss their physical meaning. 
New examples of vorticity invariants and their corresponding conserved integrals,
including new cross-helicity integrals, 
are shown. 

In Section~\ref{remarks},
we summarize all of the new vorticity-type vectors, Ertel-type scalars, and Hollmann-type cross-helicity scalars.
We also makes some concluding remarks,
including how our results can be extended the equations of inviscid fluid flow in $n>1$ dimensions. 

Finally, in Appendix~\ref{ops}, 
we review the basic algebraic and differential operations available in three-dimensional space: 
dot, cross, exterior products; grad, curl, div, Lie derivative. 
In Appendix~\ref{dictionary},
we provide a transcription between vector-calculus and differential-form notation. 
Additional mathematical background on differential forms 
and its application to fluid dynamics can be found in \Ref{ArnKhe-book,BesFri}. 

Previous work on invariants in fluid flow can be found in
\Ref{Dez,Ser,CavMor1987,CavMor1989,GusYum,KheChe,Kur,AncDar09,AncDar10,AncDarTuf,WebDasMcKHuZan,BesFri}.

\section{Types and properties of invariants}\label{prelims}

Throughout, we will work with scalars $\inv$, vectors $\vec{\inv}$, and skew tensors (bi-vectors) $\tens{\inv}=-\tens{\inv}^\t$,
since these are the most familiar types of tensorial quantities 
in mechanics and in fluid dynamics. 
Simple examples of scalars are density $\rho$ and local entropy $S$; 
vector examples are velocity $\u$ and vorticity $\vort=\vecder\times\u$; 
examples of skew tensors are 
the antisymmetric derivative of the velocity $\vecder\wedge\u = \vecder\u -(\vecder\u)^\t$ 
and the entropy-circulation tensor $\u\wedge\vecder S$.
A brief review of vectors and skew tensors,
along with dot products, wedge products, symmetric products,
as well as gradient, curl, and divergence operators, 
is provided in Appendix~\ref{ops}.

Moreover, invariants of scalar type, vector type, and skew-tensor type give rise to conserved integrals 
defined respectively on volumes, surfaces, and curves transported by the flow in a fluid.
In particular, 
helicity integral invariants and cross-helicity integral invariants
are related to scalar invariants of a certain form. 
Flux integral invariants and circulation integral invariants 
are related respectively to vector invariants and skew-tensor invariants. 

There is a simple way to transcribe everything into the setting of differential forms, 
which is explained in Appendix~\ref{dictionary}.

\subsection{Lie derivative advection and invariants }

In fluid dynamics, 
a scalar, vector, or tensor quantity, $\Inv$, 
constructed from the fluid variables and their spatial derivatives, 
is an  \emph{invariant (material quantity)} iff it satisfies $\Dt\Inv =0$ 
where
\begin{equation}\label{advLieder}
\Dt = \p_t + \Lu
\end{equation}
is the advective Lie derivative with respect to the fluid velocity $\u$. 
The Lie derivative $\Lu$ is explicitly given by 
\begin{align}
& \Lu\inv = \u\cdot\vecder\inv, 
\label{Liederscal}
\\
& \Lu \vec{\inv} = \u\cdot\vecder\vec{\inv} - \vec{\inv}\cdot\vecder\u = [\u,\vec{\inv}], 
\label{Liedervec}
\\
& \Lu \tens{\inv} = \u\cdot\vecder\tens{\inv} - \tens{\inv}\cdot\vecder\u -  (\vecder\u)^\t\cdot\tens{\inv} ,
\label{Liedertens}
\end{align}
which gives the infinitesimal change of scalars, vectors, skew tensors 
under transport along fluid streamlines.
In terms of the material derivative 
\begin{equation}\label{materialder}
\dt = \p_t + \u\cdot\vecder , 
\end{equation}
the advective Lie derivative can be expressed as
\begin{align}
& \Dt K = \dt K, 
\label{advLiederscal}
\\
& \Dt \vec{K} = \dt\vec{K} - \vec{K}\cdot\vecder\u , 
\label{advLiedervec}
\\
& \Dt \tens{K} = \dt\tens{K} - \tens{K}\cdot\vecder\u -(\vecder\u)^\t\cdot\tens{K} . 
\label{advLiedertens}
\end{align}
Recall that the material derivative describes the time derivative in a reference frame moving with the fluid.
Similarly, the advective Lie derivative describes the infinitesimal change of scalars, vectors, skew tensors in a reference frame moving with the fluid.

The (advective) Lie derivative differs from the (material) directional derivative 
by including rotation-dilation terms that take into account the tensorial nature of the quantity that it acts on. 
To explain these additional terms physically, 
consider the decomposition of the derivative of the fluid velocity into symmetric and antisymmetric tensors
\begin{equation}
\vecder\u = \tfrac{1}{2}(\symmderu + \skewderu)
\end{equation}
where
\begin{equation}\label{velocityskewder}
\skewderu =\vecder\wedge\u = \vecder\u -(\vecder\u)^\t
\end{equation}
is antisymmetric derivative of $\u$ which measures rotation of streamlines,  
and where
\begin{equation}\label{velocitysymmder}
\symmderu= \vecder\odot\u = \vecder\u +(\vecder\u)^\t 
\end{equation}
is the symmetric derivative of $\u$ which measures stretching of streamlines. 
More specifically, 
the trace of $\tfrac{1}{2}\symmderu$ given by $\vecder\cdot\u$ 
describes expansion/contraction, 
while the trace-free part of $\tfrac{1}{2}\symmderu$ describes shear. 
Then, for vectors and skew tensors, 
\begin{align}
& \dt \vec{\inv} -\Dt\vec{\inv} = \tfrac{1}{2}\vec{\inv}\cdot\symmderu + \tfrac{1}{2}\vec{\inv}\cdot\skewderu , 
\\
& \dt \tens{\inv} - \Dt\tens{\inv} = \tfrac{1}{2}(\tens{\inv}\cdot\symmderu + \symmderu\cdot\tens{\inv}) + \tfrac{1}{2}(\tens{\inv}\cdot\skewderu - \skewderu\cdot\tens{\inv}) , 
\end{align}
while for scalars, 
\begin{equation}
\dt \vec{\inv} -\Dt\vec{\inv} = 0 .
\end{equation}

A list of the basic lowest-order local invariants in fluid dynamics are shown in Table~\ref{basic-invs}. 
These invariants will be discussed further in Section~\ref{locinvs}. 
For understanding the physical meaning of the vector and skew tensor invariants, 
it is useful to recall how a vector $\vec{\inv}$ can be decomposed into 
a magnitude $|\vec{\inv}|=\sqrt{\vec{\inv}\cdot\vec{\inv}}$ 
and a direction represented by a unit vector $\hat\inv = (1/|\vec{\inv}|)\vec{\inv}$;
similarly, a skew tensor $\tens{\inv}$ can be decomposed into 
a magnitude $|\tens{\inv}|=\sqrt{\tens{\inv}\ddot\tens{\inv}}$ 
and a plane represented by a unit bi-vector $\tens{\hat\inv} = (1/|\tens{\inv}|)\tens{\inv}$. 

\begin{table}[htb]
\centering
\caption{Basic local invariants}
\label{basic-invs} 
\begin{tabular}{c|c|l|l}
\hline
Invariant & Type & Physical Meaning & Fluid system
\\
\hline\hline
$S$
& 
scalar
& 
local entropy 
& 
adiabatic
\\
$\rho$
&
scalar
&
density 
& 
incompressible
\\
$(1/\rho)\vort\cdot\vecder S$
&
scalar 
&
penetration of vortex filament
&
adiabatic
\\
&
&
into homentropic surfaces
(Ertel)
&
\\
$\vort\cdot\vecder\ln\rho$
&
scalar 
&
penetration of vortex filament
&
incompressible
\\
&
&
into constant density surfaces
&
\\
\hline
$\vort=\vecder\times\u$ 
& 
vector
& 
vorticity 
& 
constant density 
\\
$\vecder\rho\times\vecder S$
& 
vector
& 
transversality of 
homentropic 
& 
adiabatic incompressible 
\\
&&
and constant density surfaces
&
\\
\hline
$\voltens\cdot\vecder S$
& 
skew tensor
& 
entropy gradient 
&
adiabatic
\\
$\voltens\cdot\vecder\rho$
&
skew tensor
& 
density gradient 
&
incompressible
\\
\hline
\end{tabular}
\end{table}

\subsection{Integral invariants}

In fluid dynamics, 
the most physically useful type of conserved integrals
(\emph{material conservation laws}) 
are defined on moving domains given by 
volumes $\V(t)$, surfaces $\S(t)$, and curves $\C(t)$ 
that are transported by the fluid flow. 
The points $\x(t)$ comprising a transported domain obey 
\begin{equation}
\frac{d\x(t)}{dt}=\u(t,\x(t)) .
\end{equation}
It is physically natural to consider domains that are connected and (piece-wise) smooth. 
Integrals on a moving domain involve a density 
which is a function of the fluid variables and possibly their spatial derivatives 
as well as possibly $t$ and $\x$. 

A moving volume integral has the form 
\begin{equation}\label{movingvolintegral}
\int_{\V(t)} T\dV
\end{equation}
on a transported volume $\V(t)$, 
where $T$ is a scalar density
and $\dV$ is the volume element. 
This integral \eqref{movingvolintegral} is conserved 
if its time derivative vanishes 
\begin{equation}\label{invvolintegral}
\dt\int_{\V(t)} T\dV =0
\end{equation}
when it is evaluated for all solutions of a given fluid system. 
Then $\int_{\V(t)} T\dV$ is an integral invariant (namely, a constant of motion).

It is well-known that a scalar density $T$ 
yields a conserved moving-volume integral if, and only if, 
it satisfies the advection equation 
\begin{equation}\label{consscalardens}
\Dt T + (\vecder\cdot\u) T= \rho\Dt((1/\rho)T) =0
\end{equation}
holding for all solutions of a given fluid system. 
Here $\Dt$ is the advective Lie derivative \eqref{advLieder}; 
since $T$ is a scalar, this derivative coincides with the material derivative $\dt$. 
The scalar density advection equation \eqref{consscalardens} is often called 
the Reynolds transport theorem \cite{Leal-book},
and it directly relates invariant moving volume integrals to scalar invariants.

There is a similar direct relation between invariant moving surface integrals
and vector invariants, 
and also between invariant moving curve integrals and skew-tensor invariants,
which are considered next. 

A moving surface integral has the form 
\begin{equation}\label{movingsurfintegral}
\int_{\S(t)} \vec{T}\cdot\nor\dA = \int_{\S(t)} \vec{T}\cdot\dS
\end{equation}
on a transported surface $\S(t)$, 
where $\vec{T}$ is a vector density
and $\dS=\nor\dA$ is given by the surface element $\dA$ and the unit normal vector $\nor$ for $\S(t)$. 
When the surface $\S(t)$ is a closed (namely, it has no boundary), 
the normal vector $\nor$ is usually chosen to be outward directed;
when $S(t)$ is an open surface, 
the normal vector $\nor$ can be chosen in an arbitrary but continuous fashion at each point on the interior of the surface. 
An integral \eqref{movingsurfintegral} is conserved 
if its time derivative vanishes 
\begin{equation}\label{invsurfintegral}
\dt\int_{\S(t)} \vec{T}\cdot\dS = 0
\end{equation}
when it is evaluated for all solutions of a given fluid system. 
Then $\int_{\S(t)} \vec{T}\cdot\dS$ is an integral invariant 
which describes the net flux of $\vec{T}$ through the surface $\S(t)$ in the direction $\nor$. 

It can be shown that a vector density $\vec{T}$ 
yields a conserved moving-flux integral if, and only if, 
it satisfies the advection equation 
\begin{equation}\label{consvecdens}
\Dt\vec{T} + (\vecder\cdot\u)\vec{T}= \rho\Dt((1/\rho)\vec{T}) =0
\end{equation}
holding for all solutions of a given fluid system. 
Note that here the advective Lie derivative $\Dt$ consists of 
the material derivative plus a rotation-dilation term that takes into account the vectorial nature of $\vec{T}$:
\begin{equation}
\Dt\vec{T} = \dt\vec{T} - \tfrac{1}{2}\vec{T}\cdot\symmderu - \tfrac{1}{2}\vec{T}\cdot\skewderu, 
\end{equation}
where $\symmderu$ and $\skewderu$ are the symmetric and antisymmetric parts \eqref{velocitysymmder}--\eqref{velocityskewder} of the derivative of $\u$.

Finally, a moving curve integral has the form 
\begin{equation}\label{movingcurvintegral}
\int_{\C(t)} \tens{T}\ddot\mbs\nor\dl = \int_{\C(t)} (\voltens\ddot\tens{T})\cdot\ds
\end{equation}
on a transported curve $\C(t)$, 
where $\tens{T}$ is a skew-tensor density
and $\ds=\voltens\ddot\mbs\nor\dl$ is given by the line element $\dl$ and the unit normal bi-vector $\mbs\nor$ for $\C(t)$,
where $\voltens$ denotes the volume tensor. 
(Namely, $\mbs\nor$ belongs to the normal plane at each point on the curve, 
and hence $\voltens\ddot\mbs\nor=\hat s$ is a unit tangent vector along the curve.)
This integral \eqref{movingcurvintegral} is conserved 
if its time derivative vanishes 
\begin{equation}\label{invcurvintegral}
\dt\int_{\C(t)} (\voltens\ddot\tens{T})\cdot\ds =0
\end{equation}
when it is evaluated for all solutions of a given fluid system. 
Then $\int_{\C(t)} (\voltens\ddot\tens{T})\cdot\ds$ is an integral invariant 
which describes the net circulation of $\voltens\ddot\tens{T}$ along the curve $\C(t)$ in the direction $\hat s$. 

The reason for formulating moving curve integrals in terms of skew-tensor densities, 
rather than the more common approach of using scalar or vector densities, 
is shown by the simplicity of the following condition 
relating invariant circulation integrals to invariant skew-tensors. 

A skew-tensor density $\tens{T}$ yields a conserved moving-circulation integral 
if, and only if, it satisfies the advection equation 
\begin{equation}\label{constensdens}
\Dt\tens{T} + (\vecder\cdot\u)\tens{T}= \rho\Dt((1/\rho)\tens{T}) =0
\end{equation}
holding for all solutions of a given fluid system. 
Here the advective Lie derivative $\Dt$ consists of 
the material derivative plus a rotation-dilation term that takes into account the tensorial nature of $\tens{T}$:
\begin{equation}
\Dt\tens{T} = \dt\tens{T} -\tfrac{1}{2}(\tens{T}\cdot\symmderu + \symmderu\cdot\tens{T}) -\tfrac{1}{2}(\tens{T}\cdot\skewderu - \skewderu\cdot\tens{T}) . 
\end{equation}

Further discussion of material conservation laws and related developments
from a modern viewpoint appears in \Ref{Anc-review,AncChe18a,AncChe18b}. 

\subsubsection{Helicity and cross-helicity}
Any scalar invariant of the form 
$\inv = (1/\rho)\vec\xi\cdot(\vecder\times\vec\xi)$
yields a conserved \emph{helicity integral} \cite{Mof,BerFie,MofRic,ArnKhe-book}
\begin{equation}
\dt\int_{\V(t)} \vec\xi\cdot(\vecder\times\vec\xi)\dV =0 
\end{equation}
for the vorticity filaments defined by the integral curves of $\vecder\times\vec\xi$ 
for any non-gradient vector field $\vec\xi$ in a fluid. 
Helicity integrals measure the topological self-linking (knottedness) of the vorticity filaments. 
When $(1/\rho)\vec\xi\cdot(\vecder\times\vec\xi)$ is advected in the fluid,
helicity is conserved without the need for boundary conditions 
on the curl vector field $\vecder\times\vec\xi$.

The notion of helicity of a curl vector field is known to have a generalization to the mutual linking of a pair of curl vector fields:
$\int_{\V(t)} \vec\xi\cdot(\vecder\times\vec\zeta)\dV 
= \int_{\V(t)} \vec\zeta\cdot(\vecder\times\vec\xi)\dV$
when $\vec\xi\times\vec\zeta$ is tangent to the moving boundary surface $\p\V(t)$. 
This type of moving volume integral is called a \emph{cross-helicity integral} \cite{Can,Khe}.
It is conserved if either $\inv = (1/\rho)\vec\xi\cdot(\vecder\times\vec\zeta)$
or $\inv = (1/\rho)\vec\zeta\cdot(\vecder\times\vec\xi)$
is a scalar invariant,
with their densitized difference being a total divergence 
$\vecder\cdot(\vec\xi\times\vec\zeta)$.

\section{Operations on invariants}\label{operations}

In fluid dynamics, 
the dot product and cross-product, 
as well as the gradient, divergence, and curl,
have non-trivial transport properties that are important to understand 
when these operations are applied to invariants, $\Inv$. 

We begin with the dot product and cross-product. 
For any two invariant vectors $\vec{\inv}_1$ and $\vec{\inv}_2$,
their dot product and cross-product satisfy the advection identities
\begin{align}
\Dt(\vec{\inv}_1\cdot\vec{\inv}_2) & 
= \symmderu\ddot(\vec{\inv}_1\otimes\vec{\inv}_2) ,
\label{dotprodadvect}
\\
\Dt(\vec{\inv}_1\times\vec{\inv}_2) & 
= -(\vecder\cdot\u)  \vec{\inv}_1\times\vec{\inv}_2
+ \vec{\inv}_1\times(\vec{\inv}_2\cdot\symmderu)
- \vec{\inv}_2\times(\vec{\inv}_1\cdot\symmderu) ,
\label{crossprodadvect}
\end{align}
as shown in Appendix~\ref{ops}. 
Thus, these two operations are not advected in a fluid, 
unless there is no shear and no expansion/contraction so that $\symmderu=0$
(and hence $\vecder\cdot\u=0$). 
The triple product of three invariant vectors satisfies 
\begin{equation}\label{tripleprodadvect}
\Dt(\vec{\inv}_3\cdot(\vec{\inv}_1\times\vec{\inv}_2)) 
= -(\vecder\cdot\u)\vec{\inv}_3\cdot(\vec{\inv}_1\times\vec{\inv}_2) , 
\end{equation}
which is advected only in incompressible fluids. 

Next we consider the gradient, divergence, and curl. 
As shown in Appendix~\ref{ops}, 
the gradient of any scalar invariant $\inv$ satisfies the advection identity
\begin{equation}\label{gradadvect}
\Dt(\vecder\inv) = -\symmderu\cdot\vecder\inv ,
\end{equation}
and the divergence and curl of any vector invariant $\vec{\inv}$ satisfy the advection identities
\begin{align}
\Dt(\vecder\cdot\vec{\inv}) & = \vec{\inv}\cdot\vecder(\vecder\cdot\u),
\label{divadvect}
\\
\Dt(\vecder\times\vec{\inv}) & = \vecder\times(\vec{\inv}\cdot\symmderu) - (\vecder\cdot\u)\vecder\times\vec{\inv} . 
\label{curladvect}
\end{align}
Thus, the gradient and curl operations are not advected in a fluid,
due to the presence of shear and expansion/contraction,
while the divergence operation is advected only in incompressible fluids. 

All of the preceding advection properties arise from how the underlying metric, $g$, 
and volume tensor, $\voltens$, in Euclidean space $\Rnum^3$ 
behave under transport along fluid streamlines. 
The metric is a covariant symmetric tensor that directly defines the dot product,
while the volume tensor is a totally-antisymmetric contravariant tensor 
that defines the cross-product in combination with the metric, 
as explained in Appendix~\ref{ops}.
Under transport along streamlines, 
they obey 
\begin{equation}\label{metrtransport}
\lieder{\u} \invmetr = -\symmderu 
\end{equation}
where $\invmetr$ denotes the contravariant metric tensor,
and 
\begin{equation}\label{voltransport}
\lieder{\u}\, \voltens = -(\vecder\cdot\u)\voltens , 
\quad
\lieder{\u}\, \volform = (\vecder\cdot\u) \volform
\end{equation}
where $\volform$ is a 3-form that is the covariant counterpart of $\voltens$ 
and represents the volume element $dV$. 
These properties \eqref{metrtransport}--\eqref{voltransport} 
are a geometrical version of the statements 
that $\symmderu$ physically measures stretching (shear and expansion/contraction) of fluid elements, 
and that $\vecder\cdot\u$ physically measures expansion/contraction of fluid elements.

\subsection{Material operations}

We will now formulate a complete set of algebraic and differential \emph{material operations} 
that take invariants into invariants for inviscid fluid flow in three dimensions.
Specifically, any local operator acting on invariants can be expressed as a composition of the operations in this set. 
All of these operations will be presented in both geometrical and component forms. 

To begin, we observe that the volume tensor $\voltens$ has the advection property
\begin{equation}\label{voladvect}
\Dt \voltens = -(\vecder\cdot\u)\voltens . 
\end{equation}
Thus, $\voltens$ is advected in incompressible fluid flow. 
In compressible flows, the expansion/contraction of $\voltens$ 
can be compensated by noting that the fluid density $\rho$ has the same advection property, 
since a fluid volume element physically expands/contracts by the factor $\vecder\cdot\u$
in a compressible flow. 
Hence, the densitized volume tensor $(1/\rho)\voltens$ obeys
\begin{equation}\label{volinv}
\Dt((1/\rho)\voltens) = 0 , 
\end{equation}
which holds in compressible as well as incompressible fluid flow. 

This skew-tensor $(1/\rho)\voltens$ 
along with the exterior product $\wedge$ and the vector derivative operator $\vecder$
will be main ingredients in the sequel. 

\subsection{Material algebraic operations}

We first consider scalar invariants, $\inv$. 
It is easy to see that any sum or product of scalar invariants is a scalar invariant. 
More generally, we have the following straightforward result. 

\begin{prop}\label{algscal}
If $\inv_1$ and $\inv_2$ are scalar invariants, then so is $f(\inv_1,\inv_2)$
for any differentiable function $f$. 
Moreover, if $\Inv$ is an invariant, 
then so is $f(\inv_1,\inv_2)\Inv$. 
\end{prop}
The proof of the first part follows directly from the chain rule:
$\Dt f= \parder{f}{\inv_1} \Dt\inv_1 + \parder{f}{\inv_2} \Dt\inv_2 =0$. 
The second part follows from the product rule for $\Dt$. 

Next we consider vector invariants, $\vec{\inv}$. 
Although the dot product and cross product of vector invariants 
are not invariant themselves, 
the exterior product of vector invariants is an invariant. 

\begin{prop}\label{algvec}
If $\vec{\inv}_1$, $\vec{\inv}_2$ are vector invariants, then 
$\vec{\inv}_1\wedge\vec{\inv}_2$ is an invariant skew-tensor. 
\end{prop}
The proof is simply the product rule for $\Dt$. 

The triple product of vector invariants is also not an invariant (unless the fluid is incompressible), 
but it yields an invariant when it is multiplied by the fluid density $\rho$. 

\begin{prop}\label{algvectriple}
If $\vec{\inv}_1$, $\vec{\inv}_2$, $\vec{\inv}_3$ are vector invariants, then 
$\rho(\vec{\inv}_1\times\vec{\inv}_2)\cdot\vec{\inv}_3$ is an invariant scalar. 
\end{prop}
The proof amounts to combining the advection identity \eqref{tripleprodadvect} 
and the mass continuity equation for $\rho$. 

An analogous result holds for the product of a vector invariant with a skew-tensor invariant, as well as for the product of two skew-tensors. 
These products are related to the vector triple product in the following way:
\begin{equation}\label{crossvectens}
\tens{\inv}\times\vec{\inv}= \vec{\inv}\times\tens{\inv} = 2\vec{\inv}\cdot(\vec{\inv}_1\times\vec{\inv}_2)
\quad\text{when}\quad
\tens{\inv}=\vec{\inv}_1\wedge\vec{\inv}_2
\end{equation}
and 
\begin{equation}\label{crosstens}
\tens{\inv}_1\times\tens{\inv}_2 = -\tens{\inv}_2\times\tens{\inv}_1 
= (\vec{\inv}_2\times\tens{\inv}_2)\vec{\inv}_1 - (\vec{\inv}_1\times\tens{\inv}_2)\vec{\inv}_2
\quad\text{when}\quad
\tens{\inv}_1=\vec{\inv}_1\wedge\vec{\inv}_2 
\end{equation}
Their definitions in component form for general skew tensors 
are stated in Appendix~\ref{ops}.

\begin{prop}\label{algvectens}
If $\vec{\inv}_1$ is a vector invariant
and $\tens{\inv}_2$ is a skew-tensor invariant,
then $\vec{\inv}_1\wedge\tens{\inv}_2$ is an invariant totally-antisymmetric tensor
and $\rho\vec{\inv}_1\times\tens{\inv}_2$ is an invariant scalar.
\end{prop}
The proof of the first part is simply the product rule for $\Dt$,
while the second part then follows from the identity 
$\vec{\inv}_1\wedge\tens{\inv}_2 = (\vec{\inv}_1\times\tens{\inv}_2)\voltens$
combined with the advection property \eqref{voladvect}.

\begin{prop}\label{algtens}
If $\tens{\inv}_1$ and $\tens{\inv}_2$ are skew-tensor invariants,
then $\rho\tens{\inv}_1\times\tens{\inv}_2$ is an invariant vector. 
\end{prop}
The proof is similar to that of the vector triple product. 

Note that we do not need to consider totally-antisymmetric tensors like the volume tensor 
because any such tensor can be converted into a corresponding scalar 
by contraction with the volume form. 

Propositions~\ref{algscal} to~\ref{algtens}, 
and their compositions, 
encompass all possible material algebraic operations. 
This can be demonstrated more easily by using differential forms,
as shown in Appendix~\ref{dictionary}. 

Examples of the use of these operations will be presented 
in the application to fluid dynamics in Section~\ref{locinvs}.

\subsection{Material differential operations}

A primary differential operator that takes invariants into invariants is 
the Lie derivative with respect to an invariant vector, $\vec{\inv}$.  
This result is a direct mathematical consequence of the commutator identity
\begin{equation}\label{transportcommutator}
[\Dt,\lieder{\vec{\inv}}] = \lieder{\Dt\vec{\inv}} =0.
\end{equation}
From a physical viewpoint, 
for any invariant vector $\vec{\inv}$, 
its corresponding streamlines can be regarded as representing an invariant flow, 
and consequently the infinitesimal change of scalars, vectors, skew tensors 
under transport first along these streamlines and next along the fluid streamlines
is the same as their infinitesimal change 
under transport first along the fluid streamlines and next along the streamlines of the invariant flow. 

Hence we have the following main result. 

\begin{prop}\label{difflieder}
If $\Inv$ is an invariant, 
then its Lie derivative $\lieder{\vec{\inv}}\Inv$ 
with respect to any vector invariant $\vec{\inv}$ is an invariant of the same type as $\Inv$. 
\end{prop}

The expression for a Lie derivative on scalars, vectors, and tensors 
is closely related to both the gradient and the divergence operators.  
A material version of these two operators is given by the following result. 

\begin{prop}\label{diffdivgrad}
If $\inv$, $\vec{\inv}$, $\tens{\inv}$ are invariants, 
then $(1/\rho)\voltens\cdot\vecder\inv$ is an invariant skew-tensor, 
$(1/\rho)\vecder\cdot(\rho\vec{\inv})$ is an invariant scalar, 
and 
$(1/\rho)\vecder\cdot(\rho\tens{\inv})$ is an invariant vector. 
\end{prop}
The proof is a direct computation which uses the Lie derivative formulas \eqref{Liederscal}--\eqref{Liedertens}. 
In particular, 
for the first part, 
$\Dt((1/\rho)\voltens\cdot\vecder\inv) 
= (1/\rho)\voltens\cdot\Dt(\vecder\inv) + (1/\rho)\voltens\cdot(\symmderu\cdot\vecder\inv)
= (1/\rho)\voltens\cdot\vecder\Dt\inv=0$, 
after use of the advection identities \eqref{dotprodadvect} and \eqref{gradadvect}. 
For the second part, 
$\Dt((1/\rho)\vecder\cdot(\rho\vec{\inv}))
= \Dt(\vecder\cdot\vec{\inv}) + \Dt(\vecder\ln\rho)\cdot\vec{\inv}
= \vecder\cdot(\Dt\vec{\inv}) + \vec{\inv}\cdot\vecder(\Dt\ln\rho+\vecder\cdot\u)=0$
using the advection identity \eqref{divadvect} combined with 
the mass continuity equation satisfied by $\rho$. 
The third part is similar. 

All material differential operators of first order 
are encompassed by Propositions~\ref{difflieder} and~\ref{diffdivgrad},
as can be demonstrated by using differential forms,
which is shown in Appendix~\ref{dictionary}. 

Moreover, the Lie derivative operator in Proposition~\ref{difflieder}
can be constructed from the first-order differential operators in Proposition~\ref{diffdivgrad}, 
combined with the algebraic operations in Propositions~\ref{algvec}, \ref{algvectens}, \ref{algtens}, 
as follows:
\begin{align}
\lieder{\vec{\inv}_1}\inv_2 
& =(1/\rho)\vecder\cdot(\rho\inv_2\vec{\inv_1}) -(1/\rho)\vecder\cdot(\rho\vec{\inv_1})\inv_2 ,
\\
\lieder{\vec{\inv}_1}\vec{\inv}_2
& = (1/\rho)\vecder\cdot(\rho\vec{\inv}_1\wedge\vec{\inv}_2) 
+(1/\rho)\vecder\cdot(\rho\vec{\inv}_2)\vec{\inv}_1
- (1/\rho)\vecder\cdot(\rho\vec{\inv}_1)\vec{\inv}_2 , 
\\
\lieder{\vec{\inv}_1}\tens{\inv}_2
& = (1/\rho)\vecder\cdot(\rho\vec{\inv}_1\wedge\tens{\inv}_2) 
+(1/\rho)\vecder\cdot(\rho\tens{\inv}_2)\wedge\vec{\inv_1}
-(1/\rho)\vecder\cdot(\rho\vec{\inv}_1)\tens{\inv}_2 
\end{align}

From these expressions, we obtain the following first-order material differential operators involving a vector invariant. 

\begin{cor}\label{diffvec}
If $\inv$, $\vec{\inv}$, $\tens{\inv}$ are invariants, 
then for any vector invariant $\vec{\inv}_1$, 
$(1/\rho)\vecder\cdot(\rho\inv\vec{\inv}_1)$ is an invariant scalar, 
$(1/\rho)\vecder\cdot(\rho\vec{\inv}\wedge\vec{\inv}_1)$ is an invariant vector,
and $(1/\rho)\vecder\cdot(\rho\tens{\inv}\wedge\vec{\inv}_1)$ is an invariant skew-tensor. 
\end{cor}

Each of these first-order material differential operators can be composed with itself to yield 
material differential operators of arbitrary order. 
In contrast, 
compositions of the first-order differential operators in Proposition~\ref{diffdivgrad} 
gives
\begin{align}
& (1/\rho)\vecder\cdot(\voltens\cdot\vecder\inv)
=-(1/\rho)\vecder\times\vecder\inv=0,
\\
& (1/\rho)\vecder\cdot(\vecder\cdot(\rho\tens{\inv}))
=(1/\rho)(\tfrac{1}{2}\vecder\wedge\vecder)\ddot(\rho\tens{\inv})
=0,
\\
& (1/\rho)\voltens\cdot\vecder((1/\rho)\vecder\cdot(\rho\vec{\inv}))
\neq 0 .
\end{align}
All further compositions vanish. 
Hence we have the following result. 

\begin{cor}\label{diff2ndord}
$(1/\rho)\voltens\cdot\vecder((1/\rho)\vecder\cdot(\rho\vec{\inv}))$
is an invariant skew-tensor of second order 
constructed from an invariant vector $\vec{\inv}$. 
There are no invariants of higher order constructed from a single invariant vector,
and there are no invariants of second or higher order constructed from a single invariant scalar or skew-tensor.
\end{cor}

\subsection{A generating set of material operations}

To conclude these constructions, we summarize 
a generating set of all linearly independent material algebraic operations
and material differential operations of first order,
which arise from the preceding results. 

\begin{thm}\label{materialops}
\ \\
(i)
A generating set of material operations on scalar invariants, $\inv$, consists of 
\begin{equation}\label{scalgenops}
f, 
\quad
\Inv_0\cdot\grad\!, 
\end{equation}
where 
$f(\inv)$ is any differentiable function; 
and 
$\Inv_0\cdot\grad\!\!(\inv) = \Inv_0\cdot\vecder\inv$
is a directional derivative with $\Inv_0=\vec{\inv}_0,\tens{\inv}_0,(1/\rho)\voltens$ 
being any tensorial invariant. \\
(ii) 
A generating set of material operations on vector invariants and skew-tensor invariants, $\Inv=\vec{\inv},\tens{\inv}$, is given by 
\begin{equation}\label{vectensgenops}
f(\inv_0),
\quad
\invwedge,
\quad
\invtimes,
\quad
\invdiv,
\end{equation}
where 
$f(\inv_0)\Inv$ is multiplication by a differentiable function 
with $\inv_0$ being any scalar invariant;
$\invwedge\Inv$ is the exterior product
with $\Inv_0=\vec{\inv}_0,\tens{\inv}_0$ being any vector or skew-tensor invariant;
$\invtimes(\Inv) = \invtimes\Inv$ is a densitized cross-product
with $\tens{\inv}_0$ being any skew-tensor invariant;
and 
$\invdiv\Inv = (1/\rho)\vecder\cdot(\rho\Inv)$
is a densitized divergence. 
\end{thm}

As an illustration of these material operations 
and how they can be composed to generate further material operators, 
in Table~\ref{3Dmaterialops} 
we list, firstly, 
all functionally independent scalar invariants of at most first order
that are constructed from one or two invariants,
and secondly, 
all linearly independent vector and skew-tensor invariants of at most first order 
that are constructed from one or two invariants,
up to multiplication by invariant scalar functions. 
Each invariant is shown in both geometrical notation 
and component notation (in Cartesian coordinates $x^i$). 

\begin{table}[htb]
\centering
\caption{Construction of invariants}
\label{3Dmaterialops} 
\begin{tabular}{c|l|l}
\hline
Type & Geometrical Form & Component Form 
\\
\hline\hline
Scalar
& 
$\rho\tens{\inv}_2\times\vec{\inv}_1$
& 
$\rho\volform_{ijk}\inv_2^{ij}\inv_1^k$
\\
&
$(1/\rho)\vecder\cdot(\rho\vec{\inv}_1)$
& 
$(1/\rho)\nabla_i(\rho\inv_1^i)$
\\
&
$\tens{\inv}_2\times(\vecder\cdot(\rho\tens{\inv}_1))$
&
$\volform_{ijk}\inv_2^{ij}\nabla_l(\rho\inv_1^{lk})$
\\
&
$\vec{\inv}_1\cdot\vecder\inv_2$
& $\inv_1^i\nabla_i\inv_2$
\\
\hline
Vector
& 
$\rho\tens{\inv}_1\times\tens{\inv}_2$,
&
$\rho\inv_1^{ij} \volform_{jkl}\inv_2^{kl}$
\\
& 
$(1/\rho)\vecder\cdot(\rho\tens{\inv}_1)$,
& 
$(1/\rho)\nabla_j(\rho\inv_1^{ij})$
\\
&
$\tens{\inv}_2\cdot\vecder\inv_1$,
&
$\inv_2^{ij}\nabla_j\inv_1$
\\
&
$(1/\rho)\vecder\cdot(\rho\vec{\inv}_1\wedge\vec{\inv}_2)$
&
$(1/\rho)\nabla_j(\rho(\inv_1^j\inv_2^i-\inv_2^j\inv_1^i))$
\\
\hline
Skew-Tensor
&
$\vec{\inv}_1\wedge\vec{\inv}_2$
& 
$\inv_1^i \inv_2^j -\inv_2^i \inv_1^j$
\\
&
$(1/\rho)\voltens\cdot\vecder\inv_1$
& 
$(1/\rho)\volform^{ijk}\nabla_k \inv_1$
\\
&
$(1/\rho)\vec{\inv}_2\wedge(\vecder\cdot(\rho\tens{\inv}_1))$
&
$(1/\rho)(\inv_2^i \nabla_k(\rho\inv_1^{jk}) - \inv_2^j \nabla_k(\rho\inv_1^{ik}))$
\\
& 
$(1/\rho)\vecder\cdot(\rho\vec{\inv}_1\wedge\tens{\inv}_2)$
&
$(1/\rho)\nabla_k(\rho(\inv_1^k\inv_2^{ij}+\inv_1^i\inv_2^{jk}+\inv_1^j\inv_2^{ki}))$
\end{tabular}
\end{table}

The most important aspect of Theorem~\ref{materialops} is that it provides 
a simple explicit way to generate a hierarchy of invariants starting from one or more known invariants. 
Moreover, the resulting hierarchy will be complete in the sense that no additional invariants can be generated starting from the same known invariants.

\subsection{Functional independence of invariants}

It will be useful to have a general notion of functional/linear independence for invariants. 

Two scalar invariants $\inv_1$ and $\inv_2$ are said to be functionally dependent if 
$f(\inv_1,\inv_2)=0$ holds identically for some non-constant function $f$,
and otherwise the invariants $\inv_1$ and $\inv_2$ are said to be \emph{functionally independent}. 

Two vector or skew-tensor invariants $\Inv_1$ and $\Inv_2$ are said to be linearly dependent if 
$\inv_1\Inv_1 + \inv_2\Inv_2=0$ holds for some scalar invariants $\inv_1\neq 0$ and $\inv_2\neq 0$,
and otherwise if $\inv_1\Inv_1 + \inv_2\Inv_2=0$ only holds with $\inv_1=\inv_2=0$,
then the invariants $\Inv_1$ and $\Inv_2$ are said to be \emph{linearly independent}.

\section{Hierarchies of local invariants}\label{locinvs}

The equations governing inviscid adiabatic fluid flow
consist of Euler's equation 
\begin{equation}\label{eulereqn}
\u_t + \u\cdot\vecder\u = -(1/\rho)\vecder p
\end{equation} 
for the fluid velocity $\u$, 
together with the mass continuity equation 
\begin{equation}\label{denseqn}
\rho_t + \vecder\cdot(\rho \u) = 0,
\end{equation} 
for the fluid density $\rho$, 
and the adiabatic transport equation 
\begin{equation}\label{entropyeqn}
S_t + \u\cdot\vecder S =0
\end{equation} 
for the fluid entropy $S$. 
The thermodynamic relation
\begin{equation}\label{thermorel}
dE=TdS -p d(1/\rho) 
\end{equation}
determines the internal fluid energy $E$,
where $T$ is the fluid temperature and $p$ is the fluid pressure.
Here $d$ represents a material change in a thermodynamic variable. 

When the fluid is compressible, 
the fluid pressure is specified by an \eos/, 
$p=P(\rho,S)$. 
Then the thermodynamic relation \eqref{thermorel} yields 
\begin{equation}\label{ener}
E - E_0(S) =\int (1/\rho^2)P(\rho,S)d\rho = e(\rho,S), 
\end{equation}
and 
\begin{equation}\label{temp}
T=\int (1/\rho^2)\p_S P(\rho,S)d\rho +T_0(S)
\end{equation}
in terms of $\rho$ and $S$. 

If the pressure depends only on the density, $p=P(\rho)$,
then the fluid flow is \emph{barotropic}. 
In this case, the internal energy 
\begin{equation}\label{baroener}
E-E_0(S) = \int (1/\rho^2)P(\rho)d\rho = e(\rho)
\end{equation}
also depends only on $\rho$,
while the temperature
\begin{equation}\label{barotemp}
T=T_0(S) = E_0'(S)
\end{equation}
depends only on $S$. 

When the fluid is incompressible
\begin{equation}\label{incompr}
\vecder\cdot\u = 0,
\end{equation}
the density satisfies the transport equation
\begin{equation}
\rho_t+\u\cdot\vecder\rho=0,
\end{equation}
while the fluid pressure $p$ satisfies a Laplacian equation 
$\vecder\cdot((1/\rho)\vecder p) = -(\vecder\u)\ddot(\vecder\u)^\t
= \tfrac{1}{4}(|\skewderu|^2 -|\symmderu|^2)$
arising from the divergence of the Euler equation \eqref{eulereqn}. 
The thermodynamic relation \eqref{thermorel} then shows $E=E_0(S)$ and $e=0$, 
which can be viewed as a special case of the energy expression \eqref{baroener}
due to $d\rho=0$
(since $\dt\rho=0$ in incompressible flows). 
In a similar way, $T=T_0(S)$. 
In this case, both $T$ and $E$ satisfy the same transport equation as $S$. 

A fluid is \emph{homentropic} (sometimes also called isentropic) 
if the entropy $S$ is constant throughout the fluid;
a fluid is \emph{isothermal} if the temperature $T$ is constant throughout the fluid.
A fluid is \emph{adiabatic} if there is no heat transfer,
whereby $T$ is a function only of $S$ and $\rho$. 

In both cases of incompressible and compressible fluids, 
the vorticity vector 
\begin{equation}\label{vort}
\vort=\vecder\times\u
\end{equation}
satisfies the dynamical equation
\begin{equation}\label{vorteqn}
\vort_t + \u\cdot\vecder\vort - \vort\cdot\vecder\u = -(\vecder\cdot\u)\vort -\vecder(1/\rho)\times\vecder p
\end{equation} 
obtained from the curl of the Euler equation. 
At each point in the fluid, 
the vorticity vector physically describes 
the local circulation of the fluid around an infinitesimal loop 
in the plane orthogonal to this vector. 
Vortex filaments are the integral curves of the vorticity vector. 
Because the vorticity vector is divergence free
\begin{equation}\label{vortdiv}
\vecder\cdot\vort =0,
\end{equation} 
a vortex filament never terminates in the fluid 
(except at a physical boundary if a fluid is confined to a finite domain).

There are three very useful alternative forms for both the fluid velocity equation and the vorticity equation. 

First, the vector calculus identity
\begin{equation}\label{skewderuident}
\u\cdot\vecder\u -\tfrac{1}{2}\vecder(|\u|^2) = \u\cdot\skewderu = \vort\times\u
\end{equation}
leads directly to 
\begin{equation}\label{croccoeqn}
\u_t + \vort\times\u = -\tfrac{1}{2}\vecder(|\u|^2) -(1/\rho)\vecder p 
\end{equation} 
which yields
\begin{equation}\label{vorteqncurlform}
\vort_t + \vecder\times(\vort\times\u + (1/\rho)\vecder p) =0 .
\end{equation} 
This form \eqref{croccoeqn} of the fluid velocity equation is known as Crocco's theorem \cite{Cro}. 

Next, 
\begin{equation}\label{barogradp}
(1/\rho)\vecder p
= \vecder h -T\vecder S
\end{equation}
is essentially a rearrangement of the thermodynamic relation \eqref{thermorel},
where
\begin{equation}\label{enthalpy}
h= E+p/\rho
\end{equation}
is the enthalpy given by the sum of
the internal energy density $E$ and pressure-flow energy density $p/\rho$. 
This yields 
\begin{equation}\label{ueqnthermo}
\u_t + \u\cdot\vecder\u = -\vecder h +T\vecder S
\end{equation} 
and, hence,
\begin{equation}
\vort_t + \vecder\times(\vort\times\u) =\vecder T\times\vecder S . 
\end{equation} 
Note that, apart from the source term $\vecder T\times\vecder S$, 
this form of the vorticity equation is analogous to Faraday's equation in barotropic MHD. 

Last, through the vector calculus identity 
$\vecder\times(\vort\times\u) =\u\cdot\vecder\vort - \vort\cdot\vecder\u +(\vecder\cdot\u)\vort$, 
the vorticity equation \eqref{vorteqncurlform} can be expressed as
\begin{equation}\label{vorteqnadvect}
\vort_t + \u\cdot\vecder\vort - \vort\cdot\vecder\u 
= -(\vecder\cdot\u)\vort - \vecder(1/\rho)\times\vecder p , 
\end{equation} 
or alternatively in the thermodynamic form 
\begin{equation}\label{vorteqnthermo}
\vort_t + \u\cdot\vecder\vort - \vort\cdot\vecder\u 
= -(\vecder\cdot\u)\vort + \vecder T\times\vecder S 
\end{equation} 
for compressible fluids with a general \eos/.

\subsection{Basic invariants}

Inviscid fluid dynamics has several different basic local invariants, 
depending on whether the fluid flow is adiabatic or homentropic; 
compressible with a barotropic or non-barotropic \eos/; 
incompressible with constant or non-constant density.

The results in \Ref{AncDarTuf} give all local invariants of the form
$\Inv(\u,\rho,S)$ in $n>1$ dimensions.
It is straightforward in $n=3$ dimensions to extend these results to obtain
a complete classification of local invariants $\Inv(\u,\rho,S,\vecder\u,\vecder\rho,\vecder S)$. 
The basic local invariants that generate this complete lowest-order set
are shown in Table~\ref{basic-invs}. 

To begin,
we will consider each of these basic invariants
and explain the most general conditions under which each invariant holds
for inviscid fluids. 
The conserved integrals arising from these invariants 
also will be discussed.

\subsubsection{Vorticity invariant}
The basic vorticity invariant is the densitized vorticity vector
\begin{equation}\label{vortinv}
\vec{\inv}_\omega = (1/\rho)\vort  = (1/\rho)\vecder\times\u . 
\end{equation} 
To determine the conditions under which it is an invariant, 
consider the advective Lie derivative of the vorticity vector
\begin{equation}\label{vorttransport}
\Dt \vort = -(\vecder\cdot\u)\vort -\vecder(1/\rho)\times\vecder p 
\end{equation}
which is obtained from the vorticity equation \eqref{vorteqn}. 
For constant density flows, 
this transport equation \eqref{vorttransport} 
shows that the vorticity vector is advected,
since $\vecder\cdot\u=0$ and $\vecder\rho=0$. 
For non-constant density flows, 
the vorticity vector itself is no longer advected,
due to the dilational term $-(\vecder\cdot\u)\vort$
and the pressure term $-\vecder(1/\rho)\times\vecder p$. 
However, the dilational term can be compensated by 
expressing the mass continuity equation \eqref{denseqn}
in the transport form 
\begin{equation}\label{denstransport}
\Dt\rho = -(\vecder\cdot\u)\rho
\end{equation}
and combining it with the vorticity transport equation \eqref{vorttransport},
yielding $\Dt((1/\rho)\vort) = -\vecder(1/\rho)\times\vecder(p/\rho)$. 
The pressure term then vanishes if (and only if) the flow has a barotropic \eos/,
because 
$\vecder(1/\rho)\times\vecder(p/\rho) = -\rho P'(\rho)\vecder(1/\rho)\times\vecder(1/\rho)=0$,
whereby
\begin{equation}
\Dt((1/\rho)\vort) = 0.
\end{equation}
For incompressible flows, 
note that the dilational term vanishes so that 
$\Dt\vort = -\vecder(1/\rho)\times\vecder p$,
but the density term does not vanish whenever the density is not constant. 

The vorticity invariant \eqref{vortinv} yields a conserved integral 
$\dt\int_{\S(t)} \vort\cdot\dS = 0$ on moving surfaces $\S(t)$. 
Since $\vort=\vecder\times\u$ is a curl, 
this integral vanishes by Stokes' theorem if $\S(t)$ is closed. 
But if $\S(t)$ has a boundary, 
then the integral instead reduces to a moving curve integral
$\dt\oint_{\C(t)} \u\cdot\ds = 0$ 
on the closed moving boundary curve $\C(t)=\p\S(t)$. 
This yields Kelvin's circulation theorem for closed moving curves.

\subsubsection{Entropy and Ertel's invariant}
In adiabatic fluid flow, 
there are three basic local invariants:
the entropy
\begin{equation}\label{entrinv}
\inv_S = S ,
\end{equation} 
the entropy gradient 
\begin{equation}\label{entrtensinv}
\tens{\inv}_S = (1/\rho)\voltens\cdot\vecder S , 
\end{equation} 
and Ertel's invariant
\begin{equation}\label{ertlinv}
\inv_\ertl = (1/\rho)\vort\cdot\vecder S 
= (1/\rho)\vecder\cdot(S\vort) =(1/\rho)\vecder\cdot(\u\times\vecder S) . 
\end{equation} 

The entropy \eqref{entrinv} clearly is an invariant 
because it satisfies 
\begin{equation}\label{entrtransport}
\Dt S =0 
\end{equation}
without any conditions on the fluid flow. 
This invariant is a local scalar of zeroth order. 

The entropy gradient \eqref{entrtensinv} arises from applying 
the material differential operator in part (i) of Theorem~\ref{materialops} 
to the entropy,
which again does not require any conditions on the fluid flow. 
This invariant skew tensor represents a plane that is tangent to the surface of constant entropy at each point in the fluid, 
since $\vecder S\cdot\tens{\inv}_S=0$. 
Its magnitude $\sqrt{\tens{\inv}_S\ddot\tens{\inv}_S} = (\sqrt{2}/\rho)|\vecder S|$
is inversely proportional to the distance between neighboring surfaces of constant entropy.

Verifying Ertel's invariant \eqref{ertlinv} is essentially the same as 
proving the potential vorticity theorem \cite{Ert}:
\begin{equation}\label{potvorttransport}
\rho\Dt((1/\rho)\vort\cdot\vecder F(\rho,S))
= \vort\cdot\vecder \Dt F(\rho,S) -(\vecder(1/\rho)\times\vecder p)\cdot\vecder F(\rho,S)
\end{equation}
which holds for any differentiable function $F(\rho,S)$. 
Substitution of $F=S$ into equation \eqref{potvorttransport}, followed by use of $\vecder p = \p_\rho P(\rho,S)\vecder\rho + \p_S P(\rho,S)\vecder S$
holding for the general \eos/ $p=P(\rho,S)$,
we see that 
\begin{equation}
\Dt((1/\rho)\vort\cdot\vecder S)=0 . 
\end{equation}

The physical meaning of Ertel's invariant is that, 
at each point in a fluid, 
it measures the amount of penetration of a vortex filament into surfaces of constant entropy
since $(\vort\cdot\vecder S)/(|\vort||\vecder S|)$ is the alignment between 
the axis of the vortex filament and the normal direction to surfaces of constant entropy, 
while $|\vort|$ is the vortex strength
and $|\vecder S|$ is inversely proportional to the distance between neighboring constant-entropy surfaces. 
In particular, wherever $\inv_\ertl=0$, 
vortex filaments lie in surfaces of constant entropy. 

The conserved integral corresponding to Ertel's invariant \eqref{ertlinv}
is $\dt\int_{\V(t)} \vecder\cdot(\u\times\vecder S) \dV = 0$ on moving volumes $\V(t)$. 
By Gauss's theorem,
this integral reduces to a conserved moving surface integral 
$\dt\oint_{\S(t)} (\u\times\vecder S)\cdot\dS = 0$ 
on the closed moving boundary surface $\S(t)=\p\V(t)$. 
This yields a conserved entropy-circulation flux 
for closed moving surfaces. 

Both Ertel's invariant and the entropy-gradient invariant are 
local first-order invariants. 
They can be combined into an invariant function $f(\inv_S,\inv_\ertl)$,
while the skew-tensor invariant \eqref{entrtensinv} 
can be generalized by multiplication with this function 
(cf.\ Proposition~\ref{algscal}).
The resulting invariants 
\begin{equation}\label{adiabnonbaro-1stordinvs}
f(S,\inv_\ertl),
\quad
f(S,\inv_\ertl)\tens{\inv}_S
\end{equation}
comprise all local invariants of at most first order 
for inviscid adiabatic non-barotropic compressible fluid flow.

\subsubsection{Density invariants}
For incompressible non-constant density fluid flow, 
the density $\rho$ is a scalar invariant 
\begin{equation}\label{densinv}
\inv_\rho = \rho
\end{equation} 
since the mass continuity equation \eqref{denstransport} 
reduces to $\Dt\rho=0$ when $\vecder\cdot\u=0$. 
The gradient of the density then yields 
an invariant skew tensor $(1/\rho)\voltens\cdot\vecder \rho = \voltens\cdot\vecder\ln \rho$
through the material differential operator in part (i) of Theorem~\ref{materialops}. 
This invariant can be combined with the density invariant 
to obtain a simpler skew-tensor invariant
\begin{equation}\label{denstensinv}
\tens{\inv}_\rho = \voltens\cdot\vecder \rho . 
\end{equation} 
At each point in an incompressible fluid, 
the skew-tensor invariant \eqref{denstensinv}
represents a plane that is tangent to the surface of constant density,
since $\vecder\rho\cdot\tens{\inv}_\rho=0$. 
Its magnitude $\sqrt{\tens{\inv}_\rho\ddot\tens{\inv}_\rho} = \sqrt{2}\,|\vecder\rho|$
is inversely proportional to the distance between neighboring surfaces of constant density.

An additional new invariant arises from Ertel's potential vorticity theorem \eqref{potvorttransport}
by putting $F=\rho$,
which yields 
$\rho\Dt((1/\rho)\vort\cdot\vecder\rho)
= \vort\cdot\vecder \Dt\rho -(\vecder(1/\rho)\times\vecder p)\cdot\vecder\rho=0$
due to the density transport equation \eqref{denstransport}. 
Note this does not rely on the vorticity being an invariant. 
Hence we see $(1/\rho)\vort\cdot\vecder\rho$ is an invariant scalar. 
Since the density itself is an invariant, 
this shows that 
\begin{equation}\label{ertldensinv}
\inv_{\ertl'} = \vort\cdot\vecder\rho 
= \vecder\cdot(\rho\vort) =\vecder\cdot(\u\times\vecder\rho) 
\end{equation}
is an Ertel-type local scalar invariant. 
Its physical meaning measures the amount of penetration of a vortex filament into surfaces of constant density at each point in a fluid, 
since $(\vort\cdot\vecder\rho)/(|\vort||\vecder\rho|)$ is the alignment between 
the axis of the vortex filament and the normal direction to the surfaces, 
while $|\vort|$ is the vortex strength
and $|\vecder\rho|$ is inversely proportional to the distance between neighboring constant-density surfaces. 
In particular, wherever $\inv_{\ertl'}=0$, 
vortex filaments lie in constant-density surfaces. 

Similarly to Ertel's invariant \eqref{ertlinv}, 
the new density invariant \eqref{ertldensinv}
yields a moving surface integral 
$\dt\oint_{\S(t)} (\u\times\vecder \rho)\cdot\dS = 0$ 
describing a conserved density-circulation flux 
on closed moving surfaces $\S(t)$. 

Clearly, any function of the two scalar invariants $\inv_\rho$ and $\inv_{\ertl'}$ 
is also a scalar invariant,
and the product of this function with $\tens{\inv}_\rho$ is a skew tensor invariant. 
This yields all local invariants of at most first order 
\begin{equation}\label{incompr-1stordinvs}
f(\rho,\inv_{\ertl'}),
\quad
f(\rho,\inv_{\ertl'})\tens{\inv}_\rho
\end{equation}
for incompressible homentropic fluid flow.

For incompressible adiabatic fluid flow,
the density gradient invariant $\tens{\inv}_\rho$
can be combined with the entropy gradient invariant $\tens{\inv}_S$
by the material cross-product operator in part (ii) of Theorem~\ref{materialops}.
This yields 
a first-order invariant vector
\begin{equation}\label{incomprrhoentrinv}
\vec{\inv}_{S,\rho}
=\tfrac{1}{2}\tens{\inv}_\rho\times\tens{\inv}_S 
= \vecder S\times\vecder\rho . 
\end{equation}
It has the physical meaning that 
it lies in the intersection of the respective surfaces on which $S$ and $\rho$ are constant, 
and its magnitude $|\vecder\rho||\vecder S|\sqrt{1-(\vecder S\cdot\vecder\rho)^2/(|\vecder\rho||\vecder S|)^2}$ 
measures the transversality of the respective tangent planes of the surfaces
and to the inverse distance between neighboring surfaces.  
In particular, $\vec{\inv}_{S,\rho}$ vanishes 
when the two tangent planes are aligned. 

This vector invariant $\vec{\inv}_{S,\rho}$ gives rise to a conserved integral 
$\dt\int_{\S(t)} (\vecder S\times\vecder\rho)\cdot\dS = 0$ on moving surfaces $\S(t)$. 
By Stokes' theorem, this integral vanishes if $\S(t)$ is closed.
When $\S(t)$ has a boundary, 
the integral instead reduces to a moving curve integral 
on the closed moving boundary curve $\C(t)=\p\S(t)$. 
This yields a conserved circulation integral 
\begin{equation}
\oint_{\C(t)} S\vecder\rho\cdot\ds =-\oint_{\C(t)} \rho\vecder S\cdot\ds
\end{equation}
for closed moving curves.

\subsection{Higher-order invariants}

From the lowest-order invariants 
$\inv_\rho$, $\inv_{\ertl'}$, $\inv_S$, $\inv_\ertl$, $\vec{\inv}_\omega$, $\tens{\inv}_\rho$, $\tens{\inv}_S$, $\vec{\inv}_{S,\rho}$, 
we can obtain higher-order local invariants 
by applying Theorem~\ref{materialops} 
and taking into account Corollary~\ref{diff2ndord}
along with the fluid flow conditions under which each invariant holds. 
Many of these higher-order invariants are of vorticity type, 
which we will discuss. 

We will start by considering adiabatic compressible fluid flow with a non-barotropic \eos/. 
Next we will specialize to a barotropic \eos/. 
Last we will consider incompressible flow, first with constant density 
and then with non-constant density.

\subsubsection{Adiabatic compressible non-barotropic fluid flow}
The basic local invariants holding for these flows consist of 
$\inv_S$, $\inv_\ertl$, $\tens{\inv}_S$. 
Applying the material differential operators in part (i) of Theorem~\ref{materialops} 
yields 
an invariant skew-tensor 
\begin{equation}\label{ertltensinv}
\tens{\inv}_{\ertl}^{(2)}
= (1/\rho)\voltens\cdot\vecder\inv_\ertl 
\end{equation}
and an invariant vector 
\begin{equation}\label{ertlSvecinv}
\vec{\inv}_{\ertl,S}^{(2)}
=\tens{\inv}_S\cdot\vecder\inv_\ertl
=-\tens{\inv}_\ertl\cdot\vecder\inv_S
= (1/\rho)\vecder\inv_\ertl\times\vecder S .
\end{equation}
These are local second-order invariants. 
All other material operations given by Theorem~\ref{materialops}
yield trivial invariants. 
In particular, 
firstly, the material algebraic operations give 
$\rho\tens{\inv}_S\times\tens{\inv}_{\ertl}^{(2)}=2\vec{\inv}_{\ertl,S}^{(2)}$;
$\rho\vec{\inv}_{\ertl,S}^{(2)}\times\tens{\inv}_S = \vecder S\cdot(\vecder\inv_\ertl\times\vecder S)=0$;
and 
$\rho\vec{\inv}_{\ertl,S}^{(2)}\times\tens{\inv}_\ertl^{(2)} = \vecder\inv_\ertl\cdot(\vecder\inv_\ertl\times\vecder S)=0$.
Secondly, the material differential operators give 
$(1/\rho)\vecder\cdot(\rho\tens{\inv}_S)
=(1/\rho)\vecder\cdot(\voltens\cdot\vecder S) 
=(1/\rho)\vecder\times \vecder S = 0$,
$\vec{\inv}_{\ertl,S}^{(2)}\cdot\vecder\inv_S
= (1/\rho)(\vecder\inv_\ertl\times\vecder S)\cdot\vecder S=0$, 
and likewise 
$(1/\rho)\vecder\cdot(\rho\tens{\inv}_\ertl^{(2)})=0$, 
$\vec{\inv}_{\ertl,S}^{(2)}\cdot\vecder\inv_\ertl =0$. 

Since the preceding material operations are exhaustive, 
this establishes the following result. 

\begin{thm}\label{adiabnonbaro-alllocinvs}
All independent local invariants 
for inviscid adiabatic compressible fluid flow with a non-barotropic \eos/ 
are given by 
\begin{equation}\label{2ndordinvs}
S,
\quad
\inv_\ertl;
\quad
\vec{\inv}_{\ertl,S}^{(2)};
\quad
\tens{\inv}_S,
\quad
\tens{\inv}_\ertl^{(2)} .
\end{equation}
Both $\vec{\inv}_{\ertl,S}^{(2)}$ and $\tens{\inv}_\ertl^{(2)}$ are of vorticity type. 
\end{thm}

Physically, at each point in the fluid, the vorticity-type invariants 
have the following meaning:
$\tens{\inv}_\ertl^{(2)}$ represents a plane that is tangent to the surface on which $\inv_\ertl$ is constant,
while its magnitude $\sqrt{\tens{\inv}_\ertl^{(2)}\ddot\tens{\inv}_\ertl^{(2)}} = (\sqrt{2}/\rho)|\vecder\inv_\ertl|$ 
is inversely proportional to the distance between these neighboring surfaces; 
$\vec{\inv}_{\ertl,S}^{(2)}$ lies in the intersection of the respective surfaces on which $S$ and $\inv_\ertl$ are constant, 
and its magnitude $(1/\rho)|\vecder\inv_\ertl||\vecder S|\sqrt{1-(\vecder\inv_\ertl\cdot\vecder S)^2/(|\vecder\inv_\ertl||\vecder S|)^2}$ 
is proportional to the alignment between the respective tangent planes of the surfaces
and to the inverse distance between neighboring surfaces.  

The vector invariant $\vec{\inv}_{\ertl,S}^{(2)}$ gives rise to a conserved integral 
$\dt\int_{\S(t)} (\vecder\inv_\ertl\times\vecder S)\cdot\dS = 0$ on moving surfaces $\S(t)$. 
By Stokes' theorem, this integral vanishes if $\S(t)$ is closed,
since $\vecder\inv_\ertl\times\vecder S$ is a curl. 
But if $\S(t)$ has a boundary, 
then the integral reduces to a moving curve integral 
on the closed moving boundary curve $\C(t)=\p\S(t)$. 
This yields a conserved circulation integral 
\begin{equation}
\oint_{\C(t)} \inv_\ertl\vecder S\cdot\ds =-\oint_{\C(t)} S\vecder\inv_\ertl\cdot\ds
\end{equation}
for closed moving curves.

\subsubsection{Compressible barotropic fluid flow}
We will now restrict attention to compressible barotropic fluid flow, 
but with $S$ and $T(S)$ being non-constant across different fluid streamlines. 
This describes barotropic fluids in which the fluid temperature is frozen-in
and has no effect on the dynamics of the fluid velocity. 

All of the invariants \eqref{2ndordinvs} for adiabatic non-barotropic fluid flow 
are invariants in barotropic fluid flow. 
More interestingly, 
the densitized vorticity vector invariant \eqref{vortinv}
holding in barotropic fluid flow 
gives rise to further local invariants as follows. 

By applying part (i) of Theorem~\ref{materialops} 
to $\vec{\inv}_\omega$ in combination with $\inv_S$ and $\inv_\ertl$,
we reproduce Ertel's invariant
\begin{equation}\label{ertelbaroinv}
\vec{\inv}_\omega\cdot\vecder\inv_S = (1/\rho)\vort\cdot\vecder S = \inv_\ertl 
\end{equation}
and obtain another invariant scalar 
\begin{equation}\label{vortertlscalbaroinv}
\inv_{\omega,\ertl}^{(2)} 
=\vec{\inv}_\omega\cdot\vecder\inv_\ertl 
=((1/\rho)\vort\cdot\vecder)^2 S .
\end{equation}
This local second-order invariant \eqref{vortertlscalbaroinv} measures 
the amount of penetration of vortex filaments into the surfaces on which $\inv_\ertl$ is constant.

Next, using the algebraic operations in part (ii) of Theorem~\ref{materialops},
we find that no new invariants arise from 
$\vec{\inv}_\omega$ in combination with $\vec{\inv}_{\ertl,S}^{(2)}$, $\tens{\inv}_S$, and $\tens{\inv}_\ertl^{(2)}$. 
In particular:
$\vec{\inv}_\omega\wedge\vec{\inv}_{\ertl,S}^{(2)} 
= (1/\rho^2)\vort\wedge(\vecder\inv_\ertl\times\vecder S)
= \inv_\ertl \tens{\inv}_\ertl^{(2)} - \inv_{\omega,\ertl}^{(2)}\tens{\inv}_S$;
$\rho\vec{\inv}_\omega\times\tens{\inv}_S 
= (1/\rho)\vort\times(\voltens\cdot\vecder S) 
= 2\inv_\ertl $;
$\rho\vec{\inv}_\omega\times\tens{\inv}_\ertl^{(2)} 
= (1/\rho)\vort\times(\voltens\cdot\vecder\inv_\ertl) 
= 2\inv_{\omega,\ertl}^{(2)}$
(after use of some algebraic identities in Appendix~\ref{ops}).
Further, applying the differential operator in part (ii) of Theorem~\ref{materialops} to $\vec{\inv}_\omega$, 
we see that $(1/\rho)\vecder\cdot(\rho\vec{\inv}_\omega) = (1/\rho)\vecder\cdot\vort=0$ 
yields a trivial invariant. 

Hence, all local invariants up to second order 
for barotropic fluid flow are given by 
\begin{subequations}\label{2ndordbaroinvs}
\begin{gather}
f(S,\inv_\ertl,\inv_{\omega,\ertl});
\\
f(S,\inv_\ertl,\inv_{\omega,\ertl})\vec{\inv}_\omega,
\quad
f(S,\inv_\ertl,\inv_{\omega,\ertl})\vec{\inv}_{\ertl,S}^{(2)} ;
\\
f(S,\inv_\ertl,\inv_{\omega,\ertl})\tens{\inv}_S,
\quad
f(S,\inv_\ertl,\inv_{\omega,\ertl})\tens{\inv}_\ertl^{(2)} .
\end{gather}
\end{subequations}
Apart from $S$ and $\tens{\inv}_S$, these are vorticity-type invariants.

A main difference compared to the non-barotropic case, however, 
is that the additional vorticity invariant $\vec{\inv}_\omega$ 
enables the construction of higher-order invariants for barotropic fluid flow.
It is straightforward to show that all new independent invariants 
at each successive order are produced by applying part (i) of Theorem~\ref{materialops} to the all of the lower-order invariants. 
In particular, similarly to what happens at second order, 
no new invariants arise from using the algebraic operations in part (ii) of Theorem~\ref{materialops}. 

Omitting the details, 
we find that all independent local third-order invariants consist of:
two scalar invariants
\begin{gather}
\vec{\inv}_\omega\cdot\vecder\inv_{\omega,\ertl}^{(2)} 
=((1/\rho)\vort\cdot\vecder)^3 S,
\label{3rdordscalbaroinv1}
\\
\vec{\inv}_{\ertl,S}^{(2)}\cdot\vecder\inv_{\omega,\ertl}^{(2)}
=(1/\rho)\vecder( ((1/\rho)\vort\cdot\vecder)^2 S)\cdot(
\vecder((1/\rho)\vort\cdot\vecder S)\times \vecder S ) ;
\label{3rdordscalbaroinv2}
\end{gather}
two vector invariants
\begin{gather}
(1/\rho)\vecder\inv_S\times\vecder\inv_{\omega,\ertl}^{(2)}
=(1/\rho)\vecder S\times\vecder( ((1/\rho)\vort\cdot\vecder)^2 S),
\label{3rdordvecbaroinv1}
\\
(1/\rho)\vecder\inv_\ertl\times\vecder\inv_{\omega,\ertl}^{(2)}
=(1/\rho)(\vecder((1/\rho)\vort\cdot\vecder S
)\times\vecder( ((1/\rho)\vort\cdot\vecder)^2 S);
\label{3rdordvecbaroinv2}
\end{gather}
and a skew-tensor invariant
\begin{equation}
(1/\rho)\voltens\cdot\vecder\inv_{\omega,\ertl}^{(2)}
= (1/\rho)\voltens\cdot\vecder( ((1/\rho)\vort\cdot\vecder)^2 S);
\label{3rdordtensbaroinv}
\end{equation}

Going to higher orders, we have the following main result. 

\begin{thm}\label{baro-alllocinvs}
(i)
For inviscid adiabatic compressible fluid flow with a barotropic \eos/, 
all independent local invariants of order $n\geq1$ 
are recursively generated by
\begin{align}
\inv^{(n+1)} & = \vec{\inv}^{(m)}\cdot\vecder\inv^{(n)} ,
\label{scalinvrecursion}\\
\vec{\inv}^{(n+1)} & = \tens{\inv}^{(m)}\cdot\vecder\inv^{(n)} ,
\label{vecinvrecursion}\\
\tens{\inv}^{(n+1)} & = (1/\rho)\voltens\cdot\vecder\inv^{(n)} ,
\label{tensinvrecursion}
\end{align}
with $m=0,1,\ldots,n$, 
starting from 
$\inv^{(1)}=\inv_\ertl$ which is Ertel's invariant \eqref{ertlinv}, 
$\vec{\inv}^{(1)}=\vec{\inv}_\omega$ which is the vorticity invariant \eqref{vortinv}, 
and 
$\tens{\inv}^{(1)}=\tens{\inv}_S$ which is the entropy-gradient invariant \eqref{entrtensinv}.
In this hierarchy, each invariant other than $\tens{\inv}_S$ 
is of vorticity type.
(ii) 
For inviscid homentropic compressible fluid flow with a barotropic \eos/, 
the only local invariant is the vorticity \eqref{vortinv}. 
\end{thm}

\subsubsection{Constant-density fluid flow}
In these fluid flows, 
the densitized vorticity vector \eqref{vortinv} 
which is an invariant for barotropic fluid flow
is still an invariant vector. 

Consequently, 
for constant-density flows in which the entropy $S$ and temperature $T(S)$ are frozen-in,
whereby they do not affect the dynamics of the fluid velocity
but are non-constant across different fluid streamlines,
all of the invariants for adiabatic barotropic fluid flow 
shown in Theorem~\ref{baro-alllocinvs} continue to hold. 
Moreover, the density factor $1/\rho$ (being constant) can be dropped. 

In contrast, for homentropic constant-density flows,
where $S$ and $T$ are constant through the fluid, 
the basic adiabatic invariants $\inv_S$, $\inv_\ertl$, $\tens{\inv}_S$
are obviously trivial. 
As a result, the only remaining local invariant is the vorticity vector \eqref{vortinv}. 

\begin{thm}\label{constdens-alllocinvs}
(i) For inviscid adiabatic constant-density fluid flow, 
all independent local invariants of order $n\geq1$ 
are generated by the recursions 
\eqref{scalinvrecursion}, \eqref{vecinvrecursion}, \eqref{tensinvrecursion}, 
starting from 
Ertel's invariant $\inv^{(1)}=\vort\cdot\vecder S$, 
the vorticity invariant $\vec{\inv}^{(1)}=\vort$,
and the entropy-gradient invariant $\tens{\inv}^{(1)}=\voltens\cdot\vecder S$. 
(ii) For inviscid homentropic constant-density fluid flow, 
the only local invariant is the vorticity vector $\vec{\inv}^{(1)}=\vort$. 
\end{thm}

\subsubsection{Incompressible non-constant density fluid flow}
For these flows, 
the densitized vorticity vector \eqref{vortinv} is no longer an invariant. 
The basic invariants consist of 
$\inv_\rho$, $\inv_{\ertl'}$, $\tens{\inv}_\rho$. 

If the flow is also homentropic, 
then there are no other local invariants. 
But if the flow is adiabatic, 
then all of the local invariants given by Theorem~\ref{adiabnonbaro-alllocinvs}
for adiabatic non-barotropic fluid flow 
continue to hold. 
Moreover, additional local invariants arise 
by using material operations in Theorem~\ref{materialops} to combine 
these adiabatic invariants 
$S$, $\inv_\ertl$, $\vec{\inv}_{\ertl,S}^{(2)}$, $\tens{\inv}_S$, $\tens{\inv}_\ertl^{(2)}$
with the density invariants
$\inv_\rho$, $\inv_{\ertl'}$, $\tens{\inv}_\rho$
and the additional invariant $\vec{\inv}_{\rho,S}$. 

A useful observation here is that any invariant can be multiplied by $\inv_\rho=\rho$. 
Consequently, we can work with the simpler adiabatic invariants 
\begin{align}
\inv_{\tilde\ertl} & = \inv_\rho\inv_\ertl
= \vort\cdot\vecder S 
= \vecder\cdot(S\vort) =\vecder\cdot(\u\times\vecder S) ,
\label{incompr-ertlinv}
\\
\tens{\inv}_{\tilde S} & = \inv_\rho\tens{\inv}_S
=\voltens\cdot\vecder S,
\\ 
\tens{\inv}_{\tilde\ertl}^{(2)} & = \inv_\rho^2\tens{\inv}_\ertl^{(2)} +\inv_\ertl\tens{\inv}_\rho
= \voltens\cdot\vecder\inv_{\tilde\ertl} , 
\\
\vec{\inv}_{\tilde\ertl,S}^{(2)} & = \tens{\inv}_{\tilde S}\cdot\vecder\inv_{\tilde\ertl} = -\tens{\inv}_{\tilde\ertl}^{(2)}\cdot\vecder S
= \vecder\inv_{\tilde \ertl}\times\vecder S . 
\end{align}

Firstly, the material differential operators in Theorem~\ref{materialops} give 
four second-order invariant vectors
\begin{align}
\vec{\inv}_{\ertl',S}^{(2)} & 
=\tens{\inv}_{\tilde S}\cdot\vecder\inv_{\ertl'} 
= \vecder\inv_{\ertl'}\times\vecder S , 
\label{incomprvortentrvecinv}
\\
\vec{\inv}_{\ertl',\rho}^{(2)} & 
=\tens{\inv}_\rho\cdot\vecder\inv_{\ertl'} 
= \vecder\inv_{\ertl'}\times\vecder\rho ,
\label{incomprvortdensvecinv}
\\
\vec{\inv}_{\tilde\ertl,\rho}^{(2)} & 
=\tens{\inv}_\rho\cdot\vecder\inv_{\tilde\ertl} = -\tens{\inv}_{\tilde\ertl}^{(2)}\cdot\vecder\rho
= \vecder\inv_{\tilde\ertl}\times\vecder\rho , 
\label{incomprertldensvecinv}
\\
\vec{\inv}_{\ertl',\tilde\ertl}^{(2)} & 
=\tens{\inv}_{\tilde\ertl}^{(2)}\cdot\vecder\inv_{\ertl'} = \vecder\inv_{\ertl'}\times\vecder\inv_{\tilde\ertl} ;
\label{incomprertlvortdensvecinv}
\end{align}
four second-order invariant scalars 
\begin{align}
\inv_{\tilde\ertl,S,\rho}^{(2)} & 
= \vec{\inv}_{\tilde\ertl,S}^{(2)}\cdot\vecder\rho 
= -\vec{\inv}_{\tilde\ertl,\rho}^{(2)}\cdot\vecder S
= \vec{\inv}_{S,\rho}\cdot\vecder\inv_{\tilde\ertl}
= \vecder\inv_{\tilde\ertl}\cdot(\vecder S\times\vecder\rho) ,
\label{incomprertlentrdensscalinv}
\\
\inv_{\ertl',S,\rho}^{(2)} & 
= \vec{\inv}_{S,\rho}^{(2)}\cdot\vecder\inv_{\ertl'} 
= -\vec{\inv}_{\ertl',\rho}^{(2)}\cdot\vecder S
= \vec{\inv}_{\ertl',S}^{(2)}\cdot\vecder \rho
= \vecder\inv_{\ertl'}\cdot(\vecder S\times\vecder\rho) ,
\label{incomprvortentrdensscalinv}
\\
\inv_{\ertl',\ertl,S}^{(2)} & 
= \vec{\inv}_{\tilde\ertl,S}^{(2)}\cdot\vecder\inv_{\ertl'} 
= -\vec{\inv}_{\ertl',S}^{(2)}\cdot\vecder\inv_{\tilde\ertl}
= \vec{\inv}_{\ertl',\tilde\ertl}^{(2)}\cdot\vecder S
= \vecder S\cdot(\vecder\inv_{\ertl'}\times\vecder\inv_{\tilde\ertl}) , 
\label{incomprertlvortentrscalinv}
\\
\inv_{\ertl',\ertl,\rho}^{(2)} & 
= \vec{\inv}_{\tilde\ertl,\rho}^{(2)}\cdot\vecder\inv_{\ertl'} 
= -\vec{\inv}_{\ertl',\rho}^{(2)}\cdot\vecder\inv_{\tilde\ertl}
= \vec{\inv}_{\ertl',\tilde\ertl}^{(2)}\cdot\vecder \rho
= \vecder\rho\cdot(\vecder\inv_{\ertl'}\times\vecder\inv_{\tilde\ertl}) ;
\label{incomprertlvortdensscalinv}
\end{align}
and a second-order invariant skew-tensor
\begin{equation}
\tens{\inv}_{\ertl',\rho}^{(2)}
= \voltens\cdot\vecder\inv_{\ertl'} . 
\end{equation}

The physical meaning of the new vector invariants \eqref{incomprvortentrvecinv}--\eqref{incomprertlvortdensvecinv}
is similar to the meaning of the vector invariant \eqref{ertlSvecinv}. 

The four scalar invariants \eqref{incomprertlentrdensscalinv}--\eqref{incomprertlvortdensscalinv}
measure the triple alignment 
among the tangent planes of the respective surfaces on which 
$S$, $\rho$, $\inv_{\tilde\ertl}$, $\inv_{\ertl'}$ are constant. 
In particular,
if any two of the three surfaces are aligned at a point in the fluid, 
such that their normal vectors are parallel, 
then the corresponding scalar invariant vanishes. 

Each of these scalar invariants yields a conserved moving volume integral. 
Since the invariants have the form of a divergence, 
these integrals reduce to moving surface integrals by Gauss' theorem. 
This yields four conserved flux integrals 
$\dt\oint_{\S(t)} \inv_1(\vecder\inv_2\times\vecder\inv_3)\cdot\dS=0$,
where $\inv_1,\inv_2,\inv_3$ are any three of the four scalar invariants. 

The skew-tensor invariant $\tens{\inv}_{\ertl',\rho}^{(2)}$ 
physically represents a plane that is tangent to the surface on which $\inv_{\ertl'}$ is constant,
while its magnitude $\sqrt{2}|\vecder\inv_{\ertl'}|$ 
is inversely proportional to the distance between these neighboring surfaces. 

Finally, the process of generating invariants can be continued to higher orders. 
Similarly to what happens at second order, 
all new independent invariants that arise at each successive order
are produced by applying part (i) of Theorem~\ref{materialops} to the all of the lower-order invariants,
as well as using multiplication by $\inv_\rho$. 

This establishes the following result. 

\begin{thm}\label{incompr-alllocinvs}
(i) For inviscid adiabatic incompressible fluid flow with non-constant density, 
all independent local invariants of order $n\geq1$ 
are generated by the recursions
\eqref{scalinvrecursion}, \eqref{vecinvrecursion}, \eqref{tensinvrecursion}, 
starting from 
the Ertel-type invariants
$\inv^{(1)}=\vort\cdot\vecder S,\vort\cdot\vecder\rho$, 
the density-entropy surface transversality invariant
$\vec{\inv}^{(1)}=\vecder\rho\times\vecder S$, 
and the entropy-gradient and density-gradient invariants 
$\tens{\inv}^{(1)}=\voltens\cdot\vecder S,\voltens\cdot\vecder\rho$. 
(ii) For inviscid homentropic incompressible fluid flow with non-constant density, 
the only local invariants are 
$\inv=\rho,\vort\cdot\vecder\rho$, 
and $\tens{\inv}=\voltens\cdot\vecder\rho$. 
\end{thm}

\section{Hierarchies of nonlocal invariants}\label{nonlocinvs}

The starting point for deriving nonlocal invariants 
is a version of Weber's formulation \cite{Web}
of the fluid velocity equation in inviscid adiabatic compressible fluid flow. 
Weber's original formulation involves the use of Lagrangian coordinates, 
but a simpler formulation can be obtained by the use of differential forms 
\cite{App,WebDasMcKHuZan,BesFri}. 
Here we will employ an alternative version that uses only tensorial quantities:
\begin{equation}\label{weberueqn}
\u_t + \symmderu\cdot\u = \vecder(\tfrac{1}{2}|\u|^2 -E-p/\rho) +T\vecder S 
\end{equation} 
where $\symmderu$ is the symmetric derivative \eqref{velocitysymmder} of $\u$. 

This form of the fluid velocity equation arises in a similar way to the derivation of 
Crocco's theorem \eqref{croccoeqn}, 
with the important change that the vector calculus identity
\begin{equation}\label{symmderuident}
\u\cdot\vecder\u + \tfrac{1}{2}\vecder(|\u|^2) = \symmderu\cdot\u
\end{equation}
is used in place of the vorticity relation \eqref{skewderuident}. 

A connection to invariants comes from taking the dot product of
$\u_t + \symmderu\cdot\u$ with the densitized volume tensor $(1/\rho)\voltens$,
which yields 
\begin{equation}
(1/\rho)\voltens\cdot(\u_t + \symmderu\cdot\u)
= \Dt( (1/\rho)\voltens\cdot\u )
\end{equation}
by expressing the dot product in terms of the Euclidean metric $g$
and using the advection properties 
\eqref{metrtransport} for $g$ and \eqref{volinv} for $(1/\rho)\voltens$. 
Then the fluid velocity equation \eqref{weberueqn} shows that 
$(1/\rho)\voltens\cdot\u$ satisfies the transport equation 
\begin{equation}\label{webereqn}
\Dt\big( (1/\rho)\voltens\cdot\u \big)
= (1/\rho)\voltens\cdot\big( \vecder(\tfrac{1}{2}|\u|^2 -E-p/\rho) +T\vecder S \big) .
\end{equation}
We remark that this is the skew-tensor version of 
the differential-form equation \cite{App,WebDasMcKHuZan,BesFri}
$\Dt\vel = \d(\tfrac{1}{2}|\u|^2 -E-p/\rho) +T\d S$ 
where $\vel=\u\cdot\dx$ is a differential form corresponding to the fluid velocity,
as explained in Appendix~\ref{dictionary}.

After these preliminaries, 
the main idea is that now we will introduce all possible potentials that come naturally from
the gradient term and the temperature term in the velocity skew-tensor equation \eqref{webereqn}. 
These potentials first arose in the work of Weber \cite{Web}, Ertel \cite{Ert}, Rossby \cite{Ros}, and Hollmann \cite{Hol},
and they can be viewed alternatively as Clebsch variables 
which appear as Lagrange multipliers when an action principle is formulated for the fluid equations \eqref{eulereqn}, \eqref{denseqn}, \eqref{entropyeqn}
(see, e.g. \Ref{WebDasMcKHuZan}).
We will also relate these variables to a 1-form potential that has been considered in recent work \cite{BesFri} on invariants arising from
the Lagrangian formulation of the fluid velocity equation. 

To begin, 
consider a potential $\phi$ defined by the transport equation 
\begin{equation}\label{weber-pot1}
\dt\phi =\tfrac{1}{2}|\u|^2 - H(\rho,S)
\end{equation} 
where 
\begin{equation}
H(\rho,S) =E(\rho,S)+ P(\rho,S)/\rho
\end{equation} 
is the enthalpy \eqref{enthalpy}, 
with the fluid having a general \eos/ $p=P(\rho,S)$. 
This transport equation \eqref{weber-pot1} 
can be integrated to obtain $\phi$ 
along trajectories $\dt\x(t) = \u(t,\x(t))$ of infinitesimal fluid elements. 
In particular, 
\begin{equation}
\phi(t,\x(t)) = \phi(0,\x(0)) + \int_0^t \big( \tfrac{1}{2}|\u(t',\x(t'))|^2 - H(\rho(t',\x(t')),S(0,\x(0))) \big)\,dt'
\end{equation}
expresses $\phi$ as a nonlocal variable in terms of $\u$, $\rho$, and $S$. 
Note that $S(t,\x(t))=S(0,\x(0))$ since $S$ is advected by the flow. 

If we introduce the Clebsch velocity
\begin{equation}\label{v}
\v=\u-\vecder\phi
\end{equation} 
whose curl is the vorticity 
\begin{equation}
\vecder\times\v = \vecder\times\u = \vort , 
\end{equation}
then we can combine equations \eqref{weber-pot1} and \eqref{webereqn} 
to get the transport equation 
\begin{equation}\label{weberveqn}
\Dt\big( (1/\rho)\voltens\cdot\v \big) = (T/\rho)\voltens\cdot\vecder S .
\end{equation} 
The derivation uses the fact that the advective Lie derivative $\Dt$ 
coincides with the material derivative $\dt$ on scalars, 
together with the commutator identity 
\begin{equation}
[\Dt,(1/\rho)\voltens\cdot\vecder]f = 0 . 
\end{equation}

The transport equations for $(1/\rho)\voltens\cdot\v$ and $\vort$ 
give rise to interesting nonlocal invariants, 
including the Ertel-Rossby invariant (cf.~\eqref{rossinv})
and the Hollmann invariant (cf.~\eqref{hollinv}). 
Most importantly, 
the material operations shown in Theorem~\ref{materialops} 
can be applied to generate several hierarchies of additional nonlocal invariants. 
Some of the higher-order scalar invariants in these hierarchies 
describe new nonlocal cross-helicities, 
which we will discuss. 

We will begin by considering homentropic compressible fluid flow with a barotropic \eos/, 
and afterwards we will generalize the considerations to adiabatic compressible fluid flow, 
with both barotropic and non-barotropic \esos/.
Last we will consider incompressible fluid flow. 

Some final preliminary remarks are worth stating. 
Firstly, 
it is straightforward to show that equation \eqref{weberveqn} for $(1/\rho)\voltens\cdot\v$ is equivalent to a transport equation for $\v$:
\begin{equation}\label{veqn}
\Dt\v = - \symmderu\cdot\v  + T\vecder S . 
\end{equation} 
This transport equation, together with the relation $\u=\v+\vecder\phi$ and the transport equation \eqref{weber-pot1} for $\phi$, 
provides an equivalent dynamical description of the fluid velocity 
in inviscid adiabatic compressible fluid flow. 
Secondly, 
the vorticity has the transport equation
\begin{equation}\label{vorttransportthermo}
\Dt((1/\rho)\vort) = \vecder T\times\vecder S
\end{equation}
arising from the thermodynamic form of vorticity equation \eqref{vorteqnthermo} 
expressed in terms of the advective Lie derivative \eqref{advLieder}. 
Alternatively, equation \eqref{vorttransportthermo} can be obtained directly 
from the curl of equation \eqref{veqn} combined with the density transport equation \eqref{denstransport} and the commutator identities
\begin{equation}\label{Dtgradrel}
[\Dt,\vecder]f = -\symmderu\cdot\vecder f , 
\quad
[\Dt,\vecder\wedge]\vec{f} = -\symmderu\cdot\vecder\wedge\vec{f} 
\end{equation}
which hold by a straightforward computation 
employing the advection property \eqref{metrtransport} for the Euclidean metric.

Finally, $\phi$ and $\v$ have the following physical meaning 
related to the total energy density $\tfrac{1}{2}|\u|^2+H(\rho,S)$ of the fluid flow: 
We see from the transport equation \eqref{weber-pot1} that 
$\dt\phi$ is the deviation from equipartition of 
the kinetic energy density $\tfrac{1}{2}|\u|^2$ and the enthalpy energy density $H(\rho,S)$ 
in the total energy density. 
Hence, 
the Clebsch velocity \eqref{v} physically represents the part of the fluid velocity $\u$ 
that is dynamically driven by enthalpy and heat transfer 
apart from a contribution $\vecder\phi$ 
due to any deviation from equipartition of the total energy.

\subsection{Homentropic invariants}

In homentropic compressible fluid flow with a barotropic \eos/ $p=P(\rho)$, 
the fluid temperature $T$ and the entropy $S$ are constant throughout the fluid.
Consequently, 
the two main transport equations \eqref{weberveqn} and \eqref{vorttransportthermo}
simplify to the respective forms
\begin{align}
& \Dt\big( (1/\rho)\voltens\cdot\v \big) = 0, 
\\
& \Dt((1/\rho)\vort) = 0 . 
\end{align}
Hence, we see that 
\begin{equation}\label{vinv}
\tens{\inv}_v = (1/\rho)\voltens\cdot\v
\end{equation}
is a nonlocal skew-tensor invariant. 
It represents a plane that is orthogonal to the direction of $\v$ in the fluid,
while its magnitude $\sqrt{\tens{\inv}_v\ddot\tens{\inv}_v} = (\sqrt{2}/\rho)|\v|$ 
is proportional to the magnitude of $\v$. 

We can now apply the material algebraic operation in part (ii) of Theorem~\ref{materialops} 
to the invariants $\tens{\inv}_v$ and $\vec{\inv}_\omega$, 
yielding $\tens{\inv}_v\times\vec{\inv}_\omega =(2/\rho)\vort\cdot\v$. 
Hence we obtain a nonlocal scalar invariant
\begin{equation}\label{rossinv}
\inv_\ross = (1/\rho)\vort\cdot\v
\end{equation}
which is the Ertel-Rossby invariant \cite{Ros}. 
It measures the alignment between the streamlines of $\v$ and the corresponding vorticity filaments defined by $\vecder\times\v=\vort$,
and its corresponding conserved integral $\dt\int_{\V(t)} \vort\cdot\v\dV=0$
is the helicity of these filaments. 

Next we can obtain additional nonlocal invariants 
from $\tens{\inv}_v$ and $\inv_\ross$
by applying the material operations in part (i) of Theorem~\ref{materialops}.
This yields 
an invariant skew-tensor 
\begin{equation}\label{rossskewinv}
\tens{\inv}_\ross^{(2)} = (1/\rho)\voltens\cdot\vecder\inv_\ross ,
\end{equation}
and an invariant vector 
\begin{equation}\label{rossvecinv}
\vec{\inv}_{v,\ross}^{(2)}
=\tens{\inv}_v\cdot\vecder\inv_\ross
= (1/\rho)\vecder\inv_\ross\times\v ,
\end{equation}
as well as an invariant scalar
\begin{equation}\label{rossscalinv}
\inv_{\omega,\ross}^{(2)}
=\vec{\inv}_\omega\cdot\vecder\inv_\ross
= (1/\rho)\vort\cdot\vecder\inv_\ross .
\end{equation}
These are second-order invariants of vorticity type. 
The material operations given by part (ii) of Theorem~\ref{materialops}
yield no further non-trivial invariants. 
In particular, 
firstly, the material algebraic operations give 
$\rho\tens{\inv}_v\times\tens{\inv}_{\ross}^{(2)}=2\vec{\inv}_{v,\ross}^{(2)}$;
$\rho\vec{\inv}_\omega\times\tens{\inv}_{\ross}= 2 \inv_{\omega,\ross}^{(2)}$;
$\rho\vec{\inv}_{v,\ross}^{(2)}\times\tens{\inv}_v = \v\cdot(\vecder\inv_\ross\times\v)=0$;
$\rho\vec{\inv}_{v,\ross}^{(2)}\times\tens{\inv}_\ross^{(2)} = \vecder\inv_\ross\cdot(\vecder\inv_\ross\times\v)=0$;
and
$\vec{\inv}_\omega\wedge\vec{\inv}_{v,\ross}^{(2)} 
= (1/\rho^2)\vort\wedge(\vecder\inv_\ross\times\v)
= \inv_\ross\tens{\inv}_\ross^{(2)} - \inv_{\omega,\ross}^{(2)}\tens{\inv}_v$.
Secondly, the material differential operators give 
$(1/\rho)\vecder\cdot(\rho\tens{\inv}_\ross^{(2)})=0$;
and $(1/\rho)\vecder\cdot(\rho\vec{\inv}_{v,\ross}^{(2)})=\inv_{\omega,\ross}^{(2)}$. 

The three nonlocal vorticity-type invariants \eqref{rossskewinv}--\eqref{rossscalinv}
have the following physical meaning, 
which is connected to the surfaces on which the Ertel-Rossby helicity $\inv_\ross$ is constant in the fluid. 
$\tens{\inv}_\ross^{(2)}$ represents a plane that is tangent to each helicity surface, 
with $\sqrt{\tens{\inv}_\ross^{(2)}\ddot\tens{\inv}_\ross^{(2)}} = (1/\rho)|\vecder\inv_\ross|$
being inversely proportional to the distance between those neighboring surfaces;
$\inv_{\omega,\ross}^{(2)}$ measures the amount of penetration of vortex filaments into each helicity surface; 
and $\vec{\inv}_{v,\ross}^{(2)}$ lies in the intersection of each helicity surface
and the plane orthogonal to the streamlines of $\v$,
while $|\vec{\inv}_{v,\ross}^{(2)}|=(1/\rho)|\vecder\inv_\ross||\v|\sqrt{1-(\vecder\inv_\ross\cdot\v)^2/(|\vecder\inv_\ross||\v|)^2}$ 
is proportional to the alignment between the streamline plane and the helicity surface 
as well as to the inverse distance between neighboring helicity surfaces.  

The vector invariant $\vec{\inv}_{v,\ross}^{(2)}$ yields a conserved flux integral 
$\dt\int_{\S(t)} (\vecder\inv_\ross\times\v)\cdot\dS = 0$ on moving surfaces $\S(t)$. 
For closed moving surfaces, 
this conserved flux integral can be expressed as a moving volume integral by Gauss' theorem, 
which arises directly from the scalar invariant $\inv_{\omega,\ross}^{(2)}$. 

Together with the two basic invariants \eqref{vinv} and \eqref{rossinv},
the preceding vorticity-type invariants \eqref{rossskewinv}--\eqref{rossscalinv}
comprise all independent nonlocal invariants of at most second order 
for inviscid homentropic fluid flow with a barotropic \eos/. 

The process used to construct these invariants 
can be iterated to obtain a hierarchy of higher-order nonlocal invariants. 
This leads to the following main result. 

\begin{thm}\label{isenbaro-allnonlocinvs}
All independent nonlocal invariants of order $n\geq 1$
for inviscid homentropic compressible fluid flow with a barotropic \eos/
are generated by the recursions \eqref{scalinvrecursion}--\eqref{tensinvrecursion}
starting from 
$\tens{\inv}^{(0)} = \tens{\inv}_v$ 
which is the velocity invariant \eqref{vinv}, 
$\inv^{(1)}=\inv_\ross$ 
which is the Ertel-Rossby helicity invariant \eqref{rossinv}, 
and 
$\vec{\inv}^{(1)}=\vec{\inv}_\omega$ 
which is the local vorticity invariant \eqref{vortinv}. 
Other than $\tens{\inv}_v$, 
each invariant in this hierarchy is of vorticity type. 
\end{thm}

Some of the scalar invariants in this hierarchy describe new cross-helicities. 
The lowest-order example is given by 
$\inv^{(3)} = (1/\rho)\v\cdot(\vecder\inv_{\omega,\ross}^{(2)}\times\vecder\inv_{\ross})$, 
which yields the conserved integral 
\begin{equation}
\int_{\V(t)} \v\cdot(\vecder\inv_{\omega,\ross}^{(2)}\times\vecder\inv_{\ross})\dV
= \int_{\V(t)} (\inv_{\omega,\ross}^{(2)}\vecder\inv_{\ross})\cdot\vort\dV
= \int_{\V(t)} (\inv_{\ross}\vecder\inv_{\omega,\ross}^{(2)})\cdot\vort\dV
\end{equation}
on moving volumes $\V(t)$ with $\v\times\vecder\inv_{\ross}$ 
being tangent to the moving boundary surface $\p\V(t)$. 
This conserved integral is the cross-helicity of the pair of curl vector fields 
$\vort$ and $\vecder\inv_{\omega,\ross}^{(2)}\times\vecder\inv_{\ross}$.
Physically, the cross-helicity describes the mutual linking of 
the vorticity filaments given by this pair of curl vector fields in the fluid.

\subsection{Adiabatic non-barotropic invariants}

To generalize the previous results to adiabatic (non-homentropic) compressible fluid flow, 
we follow the idea in \Ref{Hol,Mob,WebDasMcKHuZan} 
by introducing another potential (Clebsch variable) $\psi$ 
defined by the transport equation
\begin{equation}\label{clebshT}
\Dt\psi= T(\rho,S)
\end{equation}
using the fluid temperature \eqref{temp}. 
This transport equation can be integrated in the same way as 
the transport equation for $\phi$, 
yielding 
\begin{equation}
\psi(t,\x(t)) = \psi(0,\x(0)) + \int_0^t T(\rho(t',\x(t')),S(0,\x(0)))\,dt'
\end{equation}
along trajectories $\dt\x(t) = \u(t,\x(t))$ of infinitesimal fluid elements. 
Consequently, 
$\psi$ represents a nonlocal variable in terms of $\u$, $\rho$, and $S$. 

Using both $\psi$ and $\phi$ to define an associated Clebsch velocity 
\begin{equation}\label{w}
\w=\u-\vecder\phi -\psi\vecder S = \v -\psi\vecder S , 
\end{equation} 
we can obtain the following transport formulation of the fluid velocity equation:
\begin{equation}\label{weqn}
\Dt\w = - \symmderu\cdot\w . 
\end{equation} 
The derivation is similar to the $\v$ equation \eqref{veqn} in case of homentropic fluid flow. 
Specifically, 
the advective Lie derivative of $\w$ is given by 
\begin{equation}\label{Dtw}
\Dt\w= \u_t -\vecder\Dt\phi -\Dt\psi\vecder S + \symmderu\cdot(\vecder\phi +\psi\vecder S)
\end{equation}
using the transport properties $\Dt\u = \u_t$ (due to $\Lu\u=0$) and $\Dt S=0$,
combined with the commutator identity \eqref{Dtgradrel}.
Next, using the respective transport equations \eqref{clebshT} and \eqref{weber-pot1} for $\psi$ and $\phi$, 
we see that equation \eqref{Dtw} becomes
\begin{equation}
\Dt\w= \u_t + \symmderu\cdot(\u-\w) -\vecder(\tfrac{1}{2}|\u|^2 - E -p/\rho) -T\vecder S .
\end{equation}
This equation simplifies through the Weber-type equation \eqref{weberueqn} for $\u$, 
directly yielding the velocity transport equation \eqref{weqn}.  

Here we are considering adiabatic compressible fluids with a general non-barotropic \eos/.
The Clebsch formulation \eqref{weqn} and the subsequent results 
can be simplified in the case of a barotropic \eos/, 
which we will consider later. 

Physically, $\psi\vecder S$ represents the dynamical contribution to $\u$ by heat transfer,
and hence the Clebsch velocity $\w$ describes the part of the fluid velocity $\u$ 
that is dynamically driven by enthalpy apart from a contribution $\vecder\phi$ 
due to any deviation from equipartition of the total energy
into kinetic and enthalpy contributions. 

The curl of $\w$ is related to the vorticity of $\u$ by 
\begin{equation}\label{wvort}
\vecder\times\w = \vecder\times(\u -\psi\vecder S)= \vort - \vecder\psi\times \vecder S ,
\end{equation}
which describes the part of the fluid vorticity that has no contribution from heat transfer. 
This relation is important because it yields a simple transport equation 
for $\vecder\times\w$. 
Taking advective Lie derivative of $(1/\rho)\vecder\times\w$ 
and using the advection properties \eqref{metrtransport} and \eqref{volinv}, 
we get 
\begin{equation}
\Dt((1/\rho)\vecder\times\w) = \Dt((1/\rho)\vort) - (1/\rho)\vecder\Dt\psi\times\vecder S
=0
\end{equation}
which follows from the transport equation \eqref{clebshT} for $\psi$. 
This shows that $(1/\rho)\vecder\times\w$ is advected in adiabatic compressible fluid flow.
Hence, 
\begin{equation}\label{adiabvortinv}
\wvort = (1/\rho)\vecder\times\w 
= (1/\rho)(\vort -\vecder\psi\times\vecder S)
\end{equation}
is a nonlocal invariant vector. 

Moreover, 
similarly to the derivation of the skew-tensor equation \eqref{webereqn} satisfied by $\u$,
here the transport equation \eqref{weqn} for $\w$ shows that 
\begin{equation}\label{winv}
\tens{\inv}_{w} = (1/\rho)\voltens\cdot\w 
\end{equation}
is a nonlocal invariant skew-tensor.
It physically represents a plane that is orthogonal to $\w$ 
at each point in the fluid. 

We remark that this invariant skew-tensor \eqref{winv}
has an alternative formulation 
as a differential form $\mbs{w}=\w\cdot\dx$ that is advected, $\Dt\mbs{w} = 0$,
as shown in Appendix~\ref{dictionary}. 
We also remark that,
along trajectories $\dt\x(t) = \u(t,\x(t))$ of infinitesimal fluid elements,
the 1-form $\mbs{w}(t,\x(t))$ is closely related to a 1-form potential
$\mbs{\gamma}(t,\x(t)) = -\int_0^t (1/\rho(t',\x(t'))\d p(t',\x(t'))\,dt'$
and an associated vorticity invariant $\d(\vel-\mbs{\gamma})$ 
derived recently in \Ref{BesFri},
where $\vel=\u\cdot\dx$ is the 1-form corresponding to the fluid velocity.
Specifically,
$\mbs{\gamma}$ arises from integration of the transport equation
$\Dt\mbs{\gamma} = -(1/\rho)\d p=T\d S -\d(E+p/\rho)$
and has the initial value $\mbs{\gamma}(0,\x(0))=\mbs{0}$. 
If we introduce a scalar potential
$\varphi(t,\x(t))= \int_0^t \tfrac{1}{2}|\u(t',\x(t'))|^2\,dt'$
related to the kinetic energy density by $\Dt\varphi = \tfrac{1}{2}|\u|^2$, 
then we can show that $\mbs{\gamma} +\d\varphi = \vel -\mbs{w}$
by the following steps. 
First, this relation can be expressed equivalently as 
$\mbs{\gamma} = \d(\phi-\varphi)+\psi\d S$
in terms of the Clebsch variables $\phi$, $\psi$, $\varphi$,
since $\mbs{w}=\vel-\d\phi-\psi\d S$ by equation \eqref{w}. 
Next, consider the 1-form $\mbs{\gamma} -\d(\phi-\varphi)-\psi\d S$. 
It is readily seen to be advected and it vanishes at $t=0$
if we choose the initial values for $\phi$ and $\psi$ to be $0$.
Thus, $\mbs{\gamma} -\d(\phi-\varphi)-\psi\d S=0$ holds for all $t\geq 0$,
from which we have $\vel -\mbs{\gamma} -\d\varphi=\mbs{w}$.
As a consequence, we see that the vorticity 2-forms
$\d(\vel -\mbs{\gamma})$ and $\d\mbs{w}$ are equal.
Therefore, the vorticity invariant $\d\mbs{w}$ is the same as
the advected vorticity 2-form $\d(\vel -\mbs{\gamma})$.
An alternative derivation of $\d(\vel -\mbs{\gamma})$,
using Lagrangian methods, is given in \Ref{BesFri}.
Finally, $\mbs{w}\wedge\d\mbs{w}$ yields an advected helicity 3-form,
which we will now derive in scalar form (cf.~\eqref{adiabhelinv})
from the nonlocal invariants \eqref{adiabvortinv} and \eqref{winv}. 

We can combine these two invariants \eqref{adiabvortinv} and \eqref{winv} 
to get a scalar invariant by applying the material algebraic operation in part (ii) of Theorem~\ref{materialops}. 
This yields
\begin{equation}
\rho\tens{\inv}_w \times\wvort = (2/\rho)\w\cdot(\vecder\times\w) ,
\end{equation}
and thus 
\begin{equation}\label{adiabhelinv}
\inv_{\varpi} = (1/\rho)\w\cdot(\vecder\times\w) = \w\cdot\wvort
\end{equation}
is a generalization of the Ertel-Rossby invariant \eqref{rossinv}
to adiabatic compressible fluid flow. 
It measures the alignment between the streamlines of $\w$ and the corresponding vorticity filaments defined by $\vecder\times\w$. 
The helicity of these filaments is given by the corresponding conserved integral
\begin{equation}
\dt\int_{\V(t)} (\vecder\times\w)\cdot\w\dV=0 . 
\end{equation}

Each of the nonlocal invariants \eqref{adiabvortinv} and \eqref{winv} 
gives rise to a conserved integral which describes a generalization of Kelvin's circulation theorem. 
In particular, 
the conserved integral given by $\tens{\inv}_w$ is a moving curve integral 
\begin{equation}
\oint_{\C(t)} \w\cdot\ds = \oint_{\C(t)} (\u-\psi\vecder S)\cdot\ds 
\end{equation}
which is the circulation of $\u-\psi\vecder S$ on closed transported curves $\C(t)$. 
This conserved circulation integral can be expressed as a moving surface integral by Stokes' theorem, 
giving $\int_{\S(t)} (\vecder\times\w)\cdot\dS$
where $\S(t)$ is any moving surface spanning the closed curve $\C(t)$ in the fluid,
namely $\p\S(t)=\C(t)$. 
Clearly, this moving surface integral arises directly from the vector invariant $\wvort$. 

To continue, by applying part (i) of Theorem~\ref{materialops} 
to the entropy invariant $S$ together with $\tens{\inv}_w$ and $\wvort$,
we further obtain 
a nonlocal vector invariant 
\begin{equation}\label{vSinv}
\tens{\inv}_w\cdot\vecder S 
=(1/\rho)\vecder S\times \w
=(1/\rho)\vecder S\times \v
= \vec{\inv}_{v,S} , 
\end{equation}
and a scalar invariant 
\begin{equation}\label{genertlinv}
\wvort\cdot\vecder S 
= (1/\rho)\vecder S\cdot(\vecder\times\w)
= (1/\rho)\vecder S\cdot(\vecder\times\u)
= \inv_\ertl
\end{equation}
which reproduces Ertel's invariant \eqref{ertlinv}. 
Here we have used the relation \eqref{w} between the Clebsch velocities $\w$ and $\v$,
which gives 
\begin{equation}\label{wminusv}
(\w-\v)\times \vecder S =0,
\quad
\vecder S\cdot\vecder\times(\w-\v) =0 .
\end{equation}

The nonlocal vector invariant $\vec{\inv}_{v,S}$ has the following physical meaning:
at each point in the fluid, 
it lies in the intersection of the plane orthogonal to the streamline of $\v$ 
and the surface of constant entropy,
while its magnitude $|\vec{\inv}_{v,S}|=(1/\rho)|\vecder S||\v|\sqrt{1-(\vecder S\cdot\v)^2/(|\vecder S|\v|)^2}$ 
is proportional to the alignment between the streamline plane and the entropy surface 
as well as to the inverse distance between neighboring entropy surfaces.  
The conserved integral arising from $\vec{\inv}_{v,S}$ is a moving surface integral 
$\int_{\S(t)} (\vecder S\times\v)\cdot\dS
= \int_{\S(t)} (\vecder S\times(\u-\vecder\phi))\cdot\dS$
describing a conserved flux. 
For closed moving surfaces, 
this integral reduces by Stokes' theorem to $\oint_{\S(t)} (\u\times\vecder S)\cdot\dS$, 
which is the conserved entropy-circulation flux arising from Ertel's invariant \eqref{ertlinv}. 

Three more nonlocal invariants now arise by applying part (i) of Theorem~\ref{materialops}
to Ertel's invariant $\inv_\ertl$ together with $\tens{\inv}_w$, $\wvort$, and $\vec{\inv}_{v,S}$. 
This yields
an invariant vector 
\begin{equation}\label{wertlvecinv}
\vec{\inv}_{w,\ertl}^{(2)}=\tens{\inv}_w\cdot\vecder\inv_\ertl
= (1/\rho)\vecder\inv_\ertl\times\w
=(1/\rho)\vecder((1/\rho)\vort\cdot\vecder S)\times\w ;
\end{equation}
and two invariant scalars
\begin{align}
\inv_{\varpi,\ertl}^{(2)} & 
=\wvort\cdot\vecder\inv_\ertl
= (1/\rho)\vecder((1/\rho)\vort\cdot\vecder S)\cdot(\vecder\times\w) , 
\label{wertlscalinv}
\\
\inv_\holl & = -\vec{\inv}_{v,S}\cdot\vecder\inv_\ertl 
= -(1/\rho)\v\cdot(\vecder\inv_\ertl\times\vecder S) .
\label{hollinv}
\end{align}
These are second-order invariants of vorticity type. 

The scalar invariant \eqref{hollinv} is Hollmann's invariant \cite{Hol}.
It is related to the triple alignment among the streamlines of $\v$ 
and the surfaces on which $S$ and $\inv_\ertl$ are respectively constant. 
The corresponding conserved integral is a moving volume integral 
$\dt\int_{\V(t)} \v\cdot(\vecder\inv_\ertl\times\vecder S)\dV=0$. 
For moving volumes $\V(t)$ with $\v\times\vecder S$ 
being tangent to the moving boundary surface $\p\V(t)$,
this conserved integral describes the cross-helicity of the pair of curl vector fields 
$\vecder\times\v$ and $\vecder S\times\vecder\inv_\ertl$:
\begin{equation}
\int_{\V(t)} \v\cdot(\vecder\inv_\ertl\times\vecder S))\dV 
= \int_{\V(t)} (\inv_\ertl\vecder S)\cdot(\vecder\times\v)\dV 
= \int_{\V(t)} (S\vecder\inv_\ertl)\cdot(\vecder\times\v)\dV . 
\end{equation}
Physically, the cross-helicity describes the mutual linking of 
the vorticity filaments given by $\vecder\times\v$ and $\vecder\inv_\ertl\times\vecder S$
in the fluid. 

Both additional invariants \eqref{wertlvecinv} and \eqref{wertlscalinv} 
are also related to the surfaces on which $\inv_\ertl$ is constant. 
The scalar invariant \eqref{wertlscalinv} measures 
the amount of penetration of the vorticity filaments of $\w$ into those surfaces. 
The vector invariant \eqref{wertlvecinv} lies in the intersection of 
each surface and the plane orthogonal to the streamline of $\w$
at each point in the fluid
and is inversely proportional to the distance between neighboring surfaces. 
It yields a conserved flux integral 
$\dt\int_{\S(t)} (\vecder((1/\rho)\vort\cdot\vecder S)\times\w)\cdot\dS = 0$ 
on moving surfaces $\S(t)$. 
For closed moving surfaces, 
this moving surface integral can be expressed as a moving volume integral by Gauss' theorem, 
which arises directly from the scalar invariant \eqref{wertlscalinv}. 

More second-order nonlocal invariants of vorticity type 
arise in a similar way from the helicity invariant $\inv_\varpi$. 
This yields
an invariant skew-tensor
\begin{equation}
\tens{\inv}_{\varpi}^{(2)} = (1/\rho)\voltens\cdot\vecder\inv_{\varpi} ;
\end{equation}
three invariant vectors
\begin{align}
\vec{\inv}_{w,\varpi}^{(2)} & =\tens{\inv}_w\cdot\vecder\inv_{\varpi}
= (1/\rho)\vecder\inv_{\varpi}\times\w , 
\label{whelvecinv}
\\
\vec{\inv}_{\varpi,\ertl}^{(2)} & =\tens{\inv}_\ertl\cdot\vecder\inv_{\varpi}
= - \tens{\inv}_w\cdot\vecder\inv_\ertl
= (1/\rho)\vecder\inv_{\varpi}\times\vecder\inv_\ertl , 
\label{hertlvecinv}
\\
\vec{\inv}_{\varpi,S}^{(2)} & =\tens{\inv}_S\cdot\vecder\inv_{\varpi}
= -\tens{\inv}_w^{(2)}\cdot\vecder S
= (1/\rho)\vecder\inv_{\varpi}\times\vecder S ;
\label{wSvecinv}
\end{align}
and four invariant scalars
\begin{align}
\inv_{\varpi}^{(2)} & =\wvort\cdot\vecder\inv_{\varpi}
= (1/\rho)(\vecder\times\w)\cdot\vecder\inv_{\varpi} ,
\label{whelscalinv}
\\
\inv_{v,S,\varpi}^{(2)} & =\vec{\inv}_{v,S}\cdot\vecder\inv_{\varpi}
= - \vec{\inv}_{w,\varpi}^{(2)}\cdot\vecder S
= (1/\rho)\v\cdot(\vecder S\times\vecder\inv_{\varpi})  ,
\label{vhSscalinv}
\\
\inv_{w,\ertl,\varpi}^{(2)} & =\vec{\inv}_{w,\ertl}^{(2)}\cdot\vecder\inv_{\varpi}
= - \vec{\inv}_{w,\varpi}^{(2)}\cdot\vecder\inv_{\ertl}
= (1/\rho)\w\cdot(\vecder\inv_{\ertl}\times\vecder\inv_{\varpi})  ,
\label{wertlhscalinv}
\\
\inv_{S,\varpi,\ertl}^{(2)} & =\vec{\inv}_{\ertl,S}^{(2)}\cdot\vecder\inv_{\varpi}
= - \vec{\inv}_{\varpi,S}^{(2)}\cdot\vecder\inv_{\ertl}
= (1/\rho)\vecder S\cdot(\vecder\inv_{\varpi}\times\vecder\inv_{\ertl})  , 
\label{whollinv}
\end{align}
where $\vec{\inv}_{\ertl,S}^{(2)}$ is the adiabatic vector invariant \eqref{ertlSvecinv}. 
(Note here we have used the relation \eqref{wminusv}.)

The physical interpretation of these nonlocal invariants is similar to the previous ones. 
In particular,
the two scalar invariants \eqref{vhSscalinv} and \eqref{wertlhscalinv}
give rise to conserved moving volume integrals 
\begin{equation}
\int_{\V(t)} \v\cdot(\vecder\times(\inv_\varpi\vecder S))\dV
= \int_{\V(t)} (\inv_\varpi\vecder S)\cdot(\vecder\times\v)\dV 
\end{equation}
and 
\begin{equation}
\int_{\V(t)} \w\cdot(\vecder\times(\inv_\varpi\vecder\inv_\ertl))\dV
= \int_{\V(t)} (\inv_\varpi\vecder\inv_\ertl)\cdot(\vecder\times\w)\dV ,
\end{equation}
which are cross-helicities of vorticity filaments given 
by $\vecder\times\v$ and $\vecder\times(\inv_\varpi\vecder S)$,
as well as by $\vecder\times\w$ and $\vecder\times(\inv_\varpi\vecder\inv_\ertl)$. 

Higher-order nonlocal invariants can be constructed 
from the preceding invariants 
by use of the material operations in part (i) of Theorem~\ref{materialops}.
In a similar way to the homentropic case considered previously, 
this leads to the following main result which generalizes 
Theorem~\ref{adiabnonbaro-alllocinvs}. 

\begin{thm}\label{adiab-allinvs}
All independent invariants (local and nonlocal) of order $n\geq 1$
for inviscid adiabatic compressible fluid with a non-barotropic \eos/ 
are given by the recursions \eqref{scalinvrecursion}--\eqref{tensinvrecursion}
starting from 
$\tens{\inv}^{(0)}=\tens{\inv}_w$ 
which is the Clebsch velocity invariant \eqref{winv}, 
$\tens{\inv}^{(1)}=\tens{\inv}_S$ 
which is the entropy-gradient invariant \eqref{entrtensinv}, 
$\vec{\inv}^{(1)}=\wvort,\vec{\inv}_{v,S}$ 
which are the adiabatic vorticity invariant \eqref{adiabvortinv} 
and the velocity-entropy alignment invariant \eqref{vSinv},
and $\inv^{(1)}=\inv_\varpi,\inv_\ertl$
which are the adiabatic helicity invariant \eqref{adiabhelinv} and Ertel's invariant \eqref{ertlinv}. 
In this hierarchy, 
each invariant other than
$\tens{\inv}_S$, $\tens{\inv}_w$, and $\vec{\inv}_{v,S}$
is of vorticity type. 
\end{thm}

\subsection{Adiabatic barotropic invariants}

We will now restrict attention to a barotropic \eos/ $p=P(\rho)$, 
where $S$ and $T(S)$ are non-constant across different fluid streamlines. 

All of the invariants from Theorem~\ref{adiab-allinvs} for non-barotropic fluid flow 
are invariants in barotropic fluid flow, 
but some of the nonlocal invariants reduce to the local invariants shown in Theorem~\ref{baro-alllocinvs}. 

To explain how this works, 
it is simplest to return to the Weber-type equation \eqref{weberueqn} for $\u$, 
and use the thermodynamic relation $T= E_S$
with $E=e(\rho)+E_0(S)$ being the internal fluid energy \eqref{baroener}.
This yields
\begin{equation}
T(S)\vecder S = E_0'(S)\vecder S = \vecder E_0(S) .
\end{equation}
Then we have 
\begin{equation}\label{baro-weber}
\u_t + \symmderu\cdot\u = \vecder(\tfrac{1}{2}|\u|^2 -E-p/\rho) +\vecder E_0
= \vecder(\tfrac{1}{2}|\u|^2 -e(\rho)-p(\rho)/\rho) .
\end{equation} 
We can now introduce a Clebsch variable through the transport equation 
\begin{equation}\label{baro-weber-pot1}
\dt\theta =\tfrac{1}{2}|\u|^2 -e(\rho)-p(\rho)/\rho
\end{equation} 
where $\theta$ is related to the previous Clebsch variables \eqref{weber-pot1} and \eqref{clebshT} by a line integral expression
\begin{equation}
\theta -\phi = \int \psi\vecder S\cdot d\x .
\end{equation} 
The fluid velocity equation \eqref{baro-weber} can again be expressed 
as a transport equation \eqref{weqn} but with 
\begin{equation}\label{barow}
\w=\u-\vecder\theta 
\end{equation} 
being the Clebsch velocity. 

The main consequence is that the curl of $\w$ now yields the local vorticity 
\begin{equation}
\vecder\times\w=\vecder\times\u= \vort . 
\end{equation} 
Hence, in Theorem~\ref{adiab-allinvs}, 
any adiabatic non-barotropic invariant that is nonlocal only through a dependence on $\vecder\times\w$ 
becomes a local invariant in the barotropic case.

\subsection{Invariants in incompressible flows}

In incompressible fluid flow, 
the Weber-type equation for $\u$ is given by 
\begin{equation}\label{incompr-weberueqn}
\u_t + \symmderu\cdot\u = \vecder(\tfrac{1}{2}|\u|^2 - p/\rho) +p\vecder(1/\rho) ,
\end{equation} 
with the density satisfying the transport equation 
\begin{equation}
\rho_t +\u\cdot\vecder\rho=0 ,
\end{equation} 
and the pressure satisfying the Laplacian equation 
$\vecder\cdot((1/\rho)\vecder p) 
= \tfrac{1}{4}(|\skewderu|^2 -|\symmderu|^2)$. 
Although the fluid velocity equation \eqref{incompr-weberueqn}
resembles the Weber-type formulation \eqref{weberueqn} for adiabatic compressible fluid flow, 
the previous previous potentials $\phi$ and $\psi$ no longer exist. 
Nevertheless, we are able to introduce analogous potentials
through the gradient term and the pressure in equation \eqref{incompr-weberueqn}. 

Consider the transport equation 
\begin{equation}\label{incompr-pot1}
\Dt\sigma =\tfrac{1}{2}|\u|^2 - p/\rho 
\end{equation} 
defining a potential $\sigma$. 
Along trajectories $\dt\x(t) = \u(t,\x(t))$ of infinitesimal fluid elements, 
this potential is a nonlocal variable 
\begin{equation}
\sigma(t,\x(t)) = \sigma(0,\x(0)) + \int_0^t \big( \tfrac{1}{2}|\u(t',\x(t'))|^2 - p(t',\x(t'))/\rho(t',\x(t') \big)\,dt'
\end{equation}
given in terms of $\u$, $p$ and $\rho$. 
Now we introduce the Clebsch-type velocity 
\begin{equation}\label{incompr-v}
\v=\u-\vecder\sigma .
\end{equation} 
Similarly to the analogous velocity \eqref{v} for adiabatic compressible fluids, 
$\v$ obeys the transport equation 
\begin{equation}\label{incompr-veqn}
\Dt\v = -\symmderu\cdot\v + p\vecder(1/\rho) .
\end{equation} 

The physical meaning of $\sigma$ and $\v$ is related to the total energy density 
$\tfrac{1}{2}|\u|^2+p/\rho$ of the fluid flow. 
Specifically, 
the transport equation \eqref{incompr-pot1} shows that 
$\Dt\sigma$ is the deviation from equipartition of 
the kinetic energy density $\tfrac{1}{2}|\u|^2$ and the pressure-flow energy density $p/\rho$
in the total energy density. 
Consequently, 
the Clebsch-type velocity \eqref{incompr-v} 
has the physical meaning of the part of the fluid velocity $\u$ 
that is dynamically driven by the pressure 
apart from a contribution $\vecder\sigma$ 
due to any deviation from equipartition of the total energy. 

We note that the curl of $\v$ yields the vorticity
\begin{equation}
\vecder\times\v = \vecder\times\u = \vort 
\end{equation}
which has the transport equation 
\begin{equation}\label{incompr-vort}
\Dt\vort = \vecder p\times\vecder(1/\rho) .
\end{equation}

To proceed, we will first look at constant-density fluids 
and afterwards consider non-constant density incompressible fluids. 

\subsubsection{Constant-density invariants}
For constant-density fluid flow, the transport equations 
\eqref{incompr-veqn} for the Clebsch velocity 
and \eqref{incompr-vort} for the vorticity 
reduce to the respective forms
\begin{align}
\Dt\v & = -\symmderu\cdot\v ,
\label{constdens-veqn}
\\
\Dt\vort &= 0 . 
\label{constdens-vorteqn}
\end{align} 

From the transport equation \eqref{constdens-veqn} for $\v$,
similarly to the derivation of the skew-tensor equation \eqref{webereqn} satisfied by $\u$,
we see that 
\begin{equation}\label{constdens-vinv}
\tens{\inv}_{\tilde v} = \voltens\cdot\v 
\end{equation}
is a nonlocal invariant skew-tensor. 
It physically represents a plane that is orthogonal to $\w$ 
at each point in the fluid. 
Both of the invariants $\tens{\inv}_{\tilde v}$ and $\vort$ 
give rise to a conserved integral 
$\dt\oint_{\C(t)} \v\cdot\ds=0$ 
which describes a generalization of Kelvin's circulation theorem, 
holding for closed transported curves $\C(t)$ in constant-density fluids. 

We can now obtain more nonlocal invariants by applying 
the material operations Theorem~\ref{materialops}. 
Note that, since the density $\rho$ is constant, 
it can be dropped in these operations. 

Combining the two invariants $\tens{\inv}_{\tilde v}$ and $\vort$ 
by use of the material algebraic operation in part (ii) of Theorem~\ref{materialops},
we get 
\begin{equation}
\tens{\inv}_{\tilde v}\times \vort
= 2\v\cdot(\vecder\times\v) .
\end{equation}
This gives a scalar invariant
\begin{equation}\label{constdens-helinv}
\inv_{\tilde\ross} = \v\cdot(\vecder\times\v) = \vort\cdot\v
\end{equation}
which is a counterpart of the Ertel-Rossby invariant \eqref{rossinv}. 
It measures the alignment between the streamlines of $\v$ and the corresponding vorticity filaments defined by $\vecder\times\v=\vort$. 
The resulting conserved moving volume integral yields the helicity of $\v$. 

For flows that are also homentropic, 
additional nonlocal invariants arise 
from the material differential operations in part (i) of Theorem~\ref{materialops}
applied to the invariants $\inv_{\tilde\ross}$, $\vort$, $\tens{\inv}_{\tilde v}$. 
This yields the following result, 
analogous to Theorem~\ref{isenbaro-allnonlocinvs}
holding for homentropic compressible flows. 

\begin{thm}\label{constdens-isentr-allnonlocinvs}
All independent nonlocal invariants of order $n\geq 1$
for inviscid homentropic constant-density fluid flow
are generated by the recursions \eqref{scalinvrecursion}--\eqref{tensinvrecursion}
starting from 
$\tens{\inv}^{(0)} = \voltens\cdot\v$, 
$\inv^{(1)}=\vort\cdot\v$, 
and 
$\vec{\inv}^{(1)}=\vort$. 
Each invariant in this hierarchy, other than $\voltens\cdot\v$,
is of vorticity type. 
\end{thm}

In contrast, 
for constant-density flows that are adiabatic, 
there are further nonlocal invariants 
which arise by using the material operations in Theorem~\ref{materialops} to combine 
$\inv_{\tilde\ross}$, $\vort$, $\tens{\inv}_{\tilde v}$
with the local adiabatic invariants 
from Theorem~\ref{constdens-alllocinvs}.
Specifically, the basic adiabatic invariants consist of 
the entropy-gradient invariant $\voltens\cdot\vecder S$,
the velocity-entropy alignment invariant 
\begin{equation}\label{incompr-vSinv}
\tens{\inv}_{\tilde v}\cdot\vecder S 
=\vecder S\times \v
= \vec{\inv}_{\tilde v,S} ,
\end{equation}
and the constant-density version \eqref{incompr-ertlinv} of Ertel's invariant. 
This leads to the following result.

\begin{thm}\label{constdens-adiab-allnonlocinvs}
All independent nonlocal invariants of order $n\geq 1$
for inviscid adiabatic constant-density fluid flow
are generated by the recursions \eqref{scalinvrecursion}--\eqref{tensinvrecursion}
starting from 
$\tens{\inv}^{(0)} = \voltens\cdot\v$, 
$\tens{\inv}^{(1)}=\voltens\cdot\vecder S$,
$\vec{\inv}^{(1)}=\vort$,
and 
$\inv^{(1)}=\vecder S\times\v,\vort\cdot\v,\vort\cdot\vecder S$.
Each invariant in this hierarchy is of vorticity type,
other than $\voltens\cdot\v$, $\voltens\cdot\vecder S$, and $\vecder S\times\v$.\end{thm}

One of the second-order scalar invariants in the hierarchy is 
a constant-density version of Hollmann's invariant \eqref{hollinv}, 
given by 
\begin{equation}\label{incompr-hollinv}
\inv_{\tilde\holl} = \v\cdot(\vecder S\times\vecder\inv_{\tilde\ertl}) . 
\end{equation}
Its corresponding conserved moving volume integral yields 
the cross-helicity of the pair of curl vector fields 
$\vecder\times\v$ and $\vecder\times(S\vecder\inv_{\tilde\ertl})$. 

\subsubsection{Density invariants}
For incompressible fluid flow in which the density is non-constant, 
neither the vorticity $\vort$ nor the Clebsch velocity skew-tensor $\tens{\inv}_v$ 
are invariants,
due to the density gradient terms occurring 
in the two respective transport equations \eqref{incompr-veqn} and \eqref{incompr-vort}.
However, the density itself is now an invariant, 
since $\Dt\rho =0$. 

We can compensate the density gradient term in the transport equation \eqref{incompr-veqn} for the Clebsch velocity $\v$ 
by introducing another Clebsch variable $\vartheta$ defined by 
\begin{equation}\label{incompr-pot2}
\Dt\vartheta =p . 
\end{equation} 
This transport equation can be integrated in the same way as 
the transport equation for $\sigma$, 
yielding 
\begin{equation}
\vartheta(t,\x(t)) = \vartheta(0,\x(0)) + \int_0^t p(t',\x(t'))\,dt'
\end{equation}
along trajectories $\dt\x(t) = \u(t,\x(t))$ of infinitesimal fluid elements. 
We use this Clebsch variable to define an associated Clebsch velocity 
\begin{equation}\label{incompr-w}
\w=\u-\vecder\sigma -\vartheta\vecder(1/\rho) = \v -\vartheta\vecder(1/\rho) 
\end{equation} 
Then the advective Lie derivative of $\w$ is given by 
\begin{equation}
\Dt\w= \Dt\v -\Dt\vartheta\vecder(1/\rho) - \vartheta\Dt\vecder(1/\rho)
= \Dt\v -p\vecder(1/\rho) +\vartheta\symmderu\cdot\vecder(1/\rho)
\end{equation}
using the transport equation  \eqref{incompr-pot2} 
and the commutator identity \eqref{Dtgradrel}.
Finally, using the transport equation \eqref{incompr-v} for $\v$,
we get 
\begin{equation}\label{incompr-weqn}
\Dt\w = - \symmderu\cdot\w . 
\end{equation} 

This transport equation \eqref{incompr-weqn} for the Clebsch velocity $\w$, 
together with the relation $\u=\w+\vecder\sigma +\vartheta\vecder(1/\rho)$
and the transport equation \eqref{incompr-pot2} for $\vartheta$, 
provides an equivalent dynamical description of the fluid velocity 
in inviscid incompressible fluid flow. 
Physically, $\w$ represents the part of the fluid velocity $\u$ 
that is not dynamically driven by the respective contributions 
$\vartheta\vecder(1/\rho)$ and $\vecder\sigma$ 
due to density gradients and deviations from equipartition of the total energy density 
$\tfrac{1}{2}|\u|^2+p/\rho$ of the fluid flow. 

The curl of $\w$ is related to the vorticity of $\u$ by 
\begin{equation}\label{incompr-wvort}
\vecder\times\w = \vecder\times(\v -\vartheta\vecder(1/\rho))
= \vort - \vecder\vartheta\times \vecder(1/\rho) .
\end{equation}
Similarly to the situation for adiabatic compressible fluid flow,
the vorticity relation \eqref{incompr-wvort}
leads to the transport equation 
\begin{equation}
\Dt(\vecder\times\w) = 0 ,
\end{equation}
showing that $\vecder\times\w$ is advected in incompressible fluid flow.

Hence, 
\begin{equation}\label{incompr-vortinv}
\vec{\inv}_{\tilde\varpi} = \vecder\times\w 
\end{equation}
is a nonlocal invariant vector. 
Physically, this invariant describes the part of the fluid vorticity that has no contribution from density gradients. 

Moreover, 
similarly to the constant-density case, 
here the transport equation \eqref{incompr-weqn} for $\w$ shows that 
\begin{equation}\label{incompr-winv}
\tens{\inv}_{\tilde w} = \voltens\cdot\w 
\end{equation}
is a nonlocal invariant skew-tensor.
It physically represents a plane that is orthogonal to $\w$ 
at each point in the fluid. 

Both nonlocal invariants \eqref{incompr-vortinv} and \eqref{incompr-winv} 
gives rise to a conserved integral 
$\dt\oint_{\C(t)} \w\cdot\ds=0$ 
which describes a generalization of Kelvin's circulation theorem, 
holding for closed transported curves $\C(t)$ in incompressible fluids with non-constant density.

Next we can combine these two nonlocal invariants \eqref{adiabvortinv} and \eqref{winv}
to get a scalar invariant by applying the material algebraic operation in part (ii) of Theorem~\ref{materialops}. 
This yields
\begin{equation}
\rho\tens{\inv}_{\tilde w} \times\vec{\inv}_{\tilde w}
= \rho\w\cdot(\vecder\times\w) .
\end{equation}
A useful observation now is that any invariant can be multiplied or divided by $\rho$. 
Thus we obtain 
\begin{equation}\label{incompr-helinv}
\inv_{\tilde \varpi} = \w\cdot(\vecder\times\w),
\end{equation}
which is a generalization of the constant-density invariant \eqref{constdens-helinv}
to incompressible fluid flows with non-constant density. 
The scalar invariant \eqref{incompr-helinv} 
measures the alignment between the streamlines of $\w$ and the corresponding vorticity filaments defined by $\vecder\times\w$. 
Its corresponding conserved integral is the helicity of these filaments. 

In the same way as for constant-density flows, 
there are additional nonlocal invariants for incompressible flows
with non-constant density. 
We have the following generalization of 
Theorems~\ref{constdens-isentr-allnonlocinvs} and~\ref{constdens-adiab-allnonlocinvs}.

\begin{thm}\label{incompr-allnonlocinvs}
For inviscid incompressible fluid flow with non-constant density, 
all independent nonlocal invariants of order $n\geq 1$
are generated by the recursions
\eqref{scalinvrecursion}--\eqref{tensinvrecursion}
starting from 
$\tens{\inv}^{(0)} =\voltens\cdot\w$,
$\tens{\inv}^{(1)} =\voltens\cdot\vecder\rho$, 
$\vec{\inv}^{(1)}=\vecder\rho\times\w,\vecder\times\w$,
and $\inv^{(1)}=\vort\cdot\vecder\rho,\w\cdot(\vecder\times\w)$
when the flow is homentropic, 
as well as 
$\tens{\inv}^{(1)}=\voltens\cdot\vecder S$,
and $\inv^{(1)}=\vort\cdot\vecder S,\vecder S\times\vecder\rho$
when the flow is adiabatic. 
Each invariant in this hierarchy is of vorticity type,
other than $\voltens\cdot\w$, $\voltens\cdot\vecder\rho$, $\voltens\cdot\vecder S$, $\vecder S\times\vecder\rho$.
\end{thm}

This hierarchy contains 
Ertel's invariant \eqref{incompr-ertlinv} and its density-type variant \eqref{ertldensinv},  
and Hollmann's invariant \eqref{incompr-hollinv},
all of which hold for general incompressible fluid flows. 
In addition, 
the hierarchy contains several new density-type variants of Hollmann's invariant:
\begin{align}
\inv_{\tilde\holl'} & = \w\cdot(\vecder\rho\times\vecder\inv_{\tilde\ertl}) , 
\label{incompr-hollinv1}
\\
\inv_{\tilde\holl''} & = \w\cdot(\vecder\rho\times\vecder\inv_{\ertl'}) ,
\label{incompr-hollinv2}
\\
\inv_{\tilde\holl'''} & = \w\cdot(\vecder S\times\vecder\inv_{\ertl'}) .
\label{incompr-hollinv3}
\end{align}
The first two of these nonlocal scalar invariants 
are related to the triple alignment among the streamlines of $\w$ 
and the respective surfaces on which $\rho$ and either $\inv_{\tilde\ertl}$ or $\inv_{\ertl'}$ 
are constant. 
The third nonlocal scalar invariant is similarly 
related to the triple alignment among the streamlines of $\w$ 
and the surfaces on which $\S$ and $\inv_{\ertl'}$ are constant. 
These invariants give rise to conserved moving volume integrals 
which yield the cross-helicity of the vorticity filaments $\vecder\times\w$
with each of the curl vector fields 
$\vecder\times(\rho\vecder\inv_{\tilde\ertl})$, 
$\vecder\times(\rho\vecder\inv_{\ertl'})$, 
$\vecder\times(S\vecder\inv_{\ertl'})$.

\section{Concluding remarks}\label{remarks}

In this paper,
we have developed a vector calculus approach to the determination of 
vorticity invariants 
in inviscid fluid flow in three dimensions. 
The main aim was to provide answers to several interesting open questions 
on advected invariants which arose in recent work \cite{WebDasMcKHuZan,TurYan,BesFri} 
using a formulation of the fluid equations based on differential forms. 

Our approach uses the more familiar and common vector-calculus formulation of 
the Eulerian fluid equations. 
Advected invariants in this formulation are naturally described by 
Lie dragging of scalars, vectors, and skew-tensors with respect to the fluid velocity 
and have the physical meaning of quantities that are frozen into the flow. 

We have constructed algebraic and differential operations that can be applied recursively
to derive a complete set of invariants starting from the basic known local and nonlocal invariants in inviscid fluid flow,
where the nonlocal invariants arise via Clebsch variables (potentials). 
Also we have explained how these invariants give rise to associated conserved integrals 
which are advected by the fluid flow. 
The basic types of invariants and conserved integrals consist of 
(i) advected scalars and corresponding conserved integrals on moving domains;
(ii) advected vectors and corresponding conserved flux integrals on moving surfaces (open or closed); 
(iii) advected skew-tensors and corresponding circulation integrals on moving curves (open or closed). 

As main results, 
infinite hierarchies of local and nonlocal invariants are obtained 
for both adiabatic fluid flow and homentropic fluid flow that 
are either incompressible, or compressible with barotropic and non-barotropic \esos/.
The hierarchies are complete in the sense that no further invariants can be generated from the basic local and nonlocal invariants in each hierarchy. 
In these hierarchies, the new invariants consist of
Ertel-type scalars, Hollmann-type cross-helicity scalars,
vorticity-type vectors, and gradient skew-tensors.
In particular, for incompressible fluid flow in which the density is non-constant 
across different fluid streamlines, 
a new variant of Ertel's invariant and several new variants of Hollmann's invariant
are derived, where the entropy gradient is replaced by the density gradient.
The physical meaning of these new invariants
and their corresponding conserved integrals 
has been discussed.

All of the new invariants are summarized in Tables~\ref{compr-invs} and~\ref{incompr-invs}, 
where $\inv$ denotes a helicity, or a cross-helicity, or an Ertel-type scalar,
obtained (recursively) from another entry in the table.
(In the language of differential forms,
a vector corresponds to a 2-form, and a skew-tensor corresponds to a 1-form,
while a scalar corresponds to a 0-form.)

\begin{table}[htb]
\centering
\caption{Invariants in inviscid compressible fluid flow}
\label{compr-invs} 
\begin{tabular}{|l||c|c|l|l}
\hline
Fluid flow & Type & Invariant & Clebsch variable
\\
\hline\hline
homentropic
& 
helicity
&
$(1/\rho)\v\cdot(\vecder\times\v)$
&
$\v=\u -\vecder\phi$
\\\cline{2-3}
&
vorticity vector
&
$(1/\rho)\vecder\times\v$
&
$\Dt\phi =\tfrac{1}{2}|\u|^2 -H$
\\
&
&
$(1/\rho)\v\times\vecder\inv$
&
$H=E(\rho,S)+ P(\rho,S)/\rho$
\\\cline{2-3}
&
Ertel-type scalar
&
$(1/\rho)(\vecder\times\v)\cdot\vecder\inv$
&
\\\cline{2-3}
&
Hollman-type
&
$\v\cdot(\vecder\inv_1\times\vecder\inv_2)$
&
\\
&
cross-helicity
&
&
\\\cline{2-3}
&
skew-tensors
&
$(1/\rho)\voltens\cdot\v$
&
\\
\hline
adiabatic
&
helicity
&
$(1/\rho)\w\cdot(\vecder\times\w)$
&
$\w=\v -\psi\vecder S$
\\\cline{2-3}
&
vorticity vector
&
$(1/\rho)\vecder\times\w$
&
$\Dt\psi= T(\rho,S)$
\\
&
&
$(1/\rho)\w\times\vecder\inv$
&
\\\cline{2-3}
&
Ertel-type scalar
&
$(1/\rho)(\vecder\times\w)\cdot\vecder\inv$
&
\\\cline{2-3}
&
Hollman-type
&
$\w\cdot(\vecder\inv_1\times\vecder\inv_2)$
&
\\
&
cross-helicity
&
&
\\\cline{2-3}
&
skew-tensors
&
$(1/\rho)\voltens\cdot\w$
&
\\
\hline
homentropic 
&
vorticity vector
&
$(1/\rho)\vecder\inv_1\times\vecder\inv_2$
&
\\\cline{2-3}
or adiabatic
&
Ertel-type scalar
&
$(1/\rho)(\vecder\inv_1\times\vecder\inv_2)\cdot\vecder\inv_3$
&
\\\cline{2-3}
&
gradient
&
$(1/\rho)\voltens\cdot\vecder\inv$
&
\\
&
skew-tensors
&
&
\\
\hline
\end{tabular}
\end{table}

\begin{table}[htb]
\centering
\caption{Invariants in inviscid incompressible fluid flow}
\label{incompr-invs} 
\begin{tabular}{|l||c|c|l|l}
\hline
Fluid flow & Type & Invariant & Clebsch variable
\\
\hline\hline
homentropic
& 
helicity
&
$\v\cdot(\vecder\times\v)$
&
$\v=\u -\vecder\sigma$
\\\cline{2-3}
&
vorticity vector
&
$\vecder\times\v$
&
$\Dt\sigma =\tfrac{1}{2}|\u|^2 - p/\rho$
\\
&
&
$\v\times\vecder\rho$
&
\\
&
&
$\v\times\vecder\inv$
&
\\\cline{2-3}
&
Ertel-type scalar
&
$(\vecder\times\v)\cdot\vecder\rho$
&
\\
&
&
$(\vecder\times\v)\cdot\vecder\inv$
&
\\
&
&
$\vecder\rho\cdot(\vecder\inv_1\times\vecder\inv_2)$
&
\\\cline{2-3}
&
Hollman-type
&
$\v\cdot(\vecder\rho\times\vecder\inv)$
&
\\
&
cross-helicity
&
$\v\cdot(\vecder\inv_1\times\vecder\inv_2)$
&
\\\cline{2-3}
&
skew-tensors
&
$\voltens\cdot\v$
&
\\
\hline
adiabatic
&
vorticity vector
&
$\w\times\vecder S$
&
$\w=\v -\vartheta\vecder(1/\rho)$
\\\cline{2-3}
&
Ertel-type scalar
&
$(\vecder\times\w)\cdot\vecder S$
&
$\Dt\vartheta =p$
\\
&
&
$(\vecder\times\w)\cdot\vecder\inv$
&
\\
&
&
$\w\cdot(\vecder\inv_1\times\vecder\inv_2)$
&
\\
&
&
$(\vecder S\times\vecder\rho)\cdot\vecder\inv$
&
\\
&
&
$\vecder S\cdot(\vecder\inv_1\times\vecder\inv_2)$
&
\\\cline{2-3}
&
Hollman-type
&
$\w\cdot(\vecder S\times\vecder\rho)$
&
\\
&
cross-helicity
&
$\w\cdot(\vecder S\times\vecder\inv)$
&
\\
&
&
$\w\cdot(\vecder\inv_1\times\vecder\inv_2)$
&
\\\cline{2-3}
&
skew-tensors
&
$\voltens\cdot\w$
&
\\
\hline
homentropic 
&
vorticity vector
&
$\vecder\inv_1\times\vecder\inv_2$
&
\\\cline{2-3}
or adiabatic
&
Ertel-type scalar
&
$(\vecder\inv_1\times\vecder\inv_2)\cdot\vecder\inv_3$
&
\\\cline{2-3}
&
gradient
&
$\voltens\cdot\vecder\inv$
&
\\
&
skew-tensors
&
&
\\
\hline
\end{tabular}
\end{table}

These results carry over to inviscid fluid flow in three-dimensional curved manifolds
by using the framework provided in the Appendices. 

In future work, 
we plan to develop a corresponding approach to find new 
helicity and cross-helicity conservation laws 
in both inviscid and viscous fluid flow. 

There are several ways to extend the ideas and methods developed in the present paper.
On the mathematical side,
it is straightforward to consider inviscid fluid flow in curved manifolds of any dimension.
This will be of interest to the topological study of hydrodynamics \cite{ArnKhe-book}.
On the physical side,
it is interesting to look at other hydrodynamical systems 
such as viscous fluids, magnetohydrodynamics, and multi-phase fluid and plasma models. 
We intend to explore these two directions in future work. 

Finally, 
an important question is what are the implications of the existence of an infinite number of conserved integral invariants in inviscid fluid flow? 
The invariants contain increasingly higher order derivatives of the basic fluid variables and the nonlocal Clebsch variables,
and thus are closely related to regularity properties of solutions of the fluid equations. 
Moreover, the higher-order helicities and other higher-order invariants of vorticity type 
are plausibly connected to finer scale features in the solutions. 
Indeed, it is also possible to speculate that the higher-order vorticity invariants 
could be relevant to detecting development of turbulence. 
Investigation of such possibilities will require a deeper understanding of 
both the physical and mathematical meaning of these invariants.

\appendix
\section{Geometric tensors and operations in three-dimensional space}\label{ops}

Three-dimensional space, $\Rnum^3$,
as well as more general spaces with curvature and/or non-trivial topology,
can be described mathematically as an oriented Riemannian manifold
whose tangent space at every point is isomorphic to $\Rnum^3$. 
Such manifolds have three fundamental geometrical structures:
a metric $\metr$, a volume 3-form $\volform$,
and a covariant derivative $\nabla$ satisfying $\nabla\metr=0$ and $\nabla\volform=0$. 

The metric $\metr$ has the geometrical meaning that, 
for any two non-collinear vectors $\vec{a}$ and $\vec{b}$, 
the inner product $\metr(\vec{a},\vec{b}) = \vec{a}\cdot\vec{b}$ 
is equal to the product of their lengths and the cosine of the angle between them. 
Similarly, the volume 3-form $\volform$ has the geometrical meaning that, 
for any three non-collinear vectors $\vec{a}$, $\vec{b}$, $\vec{c}$, 
the triple product $\volform(\vec{a},\vec{b},\vec{c}) = (\vec{a}\times\vec{b})\cdot\vec{c}$ 
is equal to the volume of the parallelepiped spanned by these vectors.
The geometrical meaning of the covariant derivative is that
$\nabla\vec{a}=0$ defines a vector field $\vec{a}$ that is parallel-transported.

Let $\e{i}$, $i=1,2,3$, denote a frame consisting of three orthonormal vectors
in the tangent space (at every point).
The manifold is flat if there exists a parallel-transported frame,
$\nabla\e{i}=0$, $i=1,2,3$.
This condition is equivalent to the vanishing of the Riemannian curvature tensor. 

In any frame,
the components of $\metr$ and $\volform$ are given by 
$\metr(\e{i},\e{j}) = \e{i}\cdot\e{j}=\delta_{ij}$ 
(Kronecker symbol)
and $\volform(\e{i},\e{j},\e{k}) = \e{i}\cdot(\e{j}\times\e{k})=\volform_{ijk}$ 
(Levi-Civita symbol).
Associated to the metric is an inverse metric tensor
$\invmetr = \delta^{ij} \e{i}\otimes\e{j}$.
Similarly, associated to the volume 3-form is a totally-antisymmetric volume tensor
$\voltens = \volform^{ijk} \e{i}\otimes\e{j}\otimes\e{k}$.
Here and hereafter, summation is assumed for any repeated index. 

Next we recall the basic dot and cross product operations on vectors and skew-tensors,
along with the exterior product operation. 
After that we review the different types of differential operations on scalars, vectors, and skew-tensors.

\subsection{Algebraic operations: dot, cross, exterior products}

The frame components of a vector $\vec{a}$ are given by 
$a^i = \metr(\vec{a},\e{i})=\vec{a}\cdot\e{i}$, $i=1,2,3$,
where $a^i\e{i}=\vec{a}$. 
Similarly, the frame components of a skew-tensor $\tens{a}$ are given by 
$a^{ij} = -a^{ji}= \metr(\tens{a},\e{i})=\tens{a}\cdot\e{i}$, $i=1,2,3$,
where $a^{ij}\e{i}\otimes\e{j}=\tens{a}$.
Any totally-antisymmetric tensor is a multiple of the volume tensor, $a\voltens$, 
where $a$ is a scalar.
Its frame components are simply $a\volform^{ijk}$. 

Given two vectors $\vec{a}$ and $\vec{b}$, 
their dot product is a scalar $c=\vec{a}\cdot\vec{b}$ 
and their cross product is a vector $\vec{c}=\vec{a}\times\vec{b}= -\vec{b}\times\vec{a}$. 
In frame components:
$\vec{a}\cdot\vec{b} = \delta_{ij}a^ib^j$
and $(\vec{a}\times\vec{b})^i = \volform^{i}{}_{jk} a^jb^k$.  

Two more operations are the exterior product yielding a skew-tensor 
$\tens{c} = \vec{a}\wedge\vec{b} = \vec{a}\otimes\vec{b} - \vec{b}\otimes\vec{a}$,
and the symmetric product yielding a symmetric tensor
$\vec{a}\odot\vec{b} = \vec{a}\otimes\vec{b} +\vec{b}\otimes\vec{a}$. 
Their frame components are given by 
$(\vec{a}\wedge\vec{b})^{ij} = a^i b^j - b^i a^j$,
and 
$(\vec{a}\odot\vec{b})^{ij} = a^i b^j + b^i a^j$. 

These products have a natural extension in which one vector is replaced by a skew-tensor. 
Given a vector $\vec{a}$ and a skew-tensor $\tens{b}$,
their dot product is a vector $\vec{c}= \vec{a}\cdot\tens{b} = - \tens{b}\cdot\vec{a}$,
which has the frame components 
$(\vec{a}\cdot\tens{b})^i = \delta_{jk}a^j b^{ki} = - \delta_{jk} b^{ij} a^k$. 
Their cross product is a scalar $c= \vec{a}\times\tens{b} = \tens{b}\times\vec{a}$
defined in frame components as $\volform_{ijk}a^i b^{jk} = \volform_{ijk}b^{ij} a^k$. 
The exterior product $\vec{a}\wedge\tens{b}$ is a totally-antisymmetric tensor 
whose frame components are given by 
$(\vec{a}\wedge\tens{b})^{ijk} = a^i b^{jk} + a^j b^{ki} + a^k b^{ij}$. 
This tensor is a scalar multiple of the volume tensor $\voltens$, in particular
$\vec{a}\wedge\tens{b} = \tfrac{1}{2}(\vec{a}\times\tens{b})\voltens$. 

There is a further natural extension to products of two skew-tensors. 
The dot product of $\tens{a}$ and $\tens{b}$ is a tensor without symmetry. 
Antisymmetrizing this tensor yields a skew-tensor 
$\tens{c}= \tens{a}\cdot\tens{b} - \tens{b}\cdot\tens{a}$,
while symmetrizing instead yields a symmetric tensor 
$\tens{a}\cdot\tens{b} + \tens{b}\cdot\tens{a}$,
where in frame components
$(\tens{a}\cdot\tens{b})^{ij} = \delta_{kl} a^{ik} b^{lj}$. 
The cross product of $\tens{a}$ and $\tens{b}$ is a vector 
$\vec{c}= \tens{a}\times\tens{b} = -\tens{b}\times\tens{a}$
whose frame components are given by 
$(\tens{a}\times\tens{b})^i = \volform_{jkl} a^{ij} b^{kl} =-\volform_{jkl} b^{ij} a^{kl}$. 
The double-dot product of $\tens{a}$ and $\tens{b}$ is a scalar 
$\tens{a}\ddot\tens{b} = \delta_{ik}\delta_{jl} a^{ij} b^{kl}$. 

Using the volume tensor, 
a vector $\vec{a}$ can be converted into a skew-tensor 
by $\tens{a}_\voltens = \voltens\cdot\vec{a}$,
whose frame components are $a_\voltens{}^{ij} = \volform^{ij}{}_{k} a^k$
with $\volform^{ij}{}_{k} = \volform^{ijl}\delta_{lk}$. 
Conversely, 
a skew-tensor $\tens{a}$ can be converted into a vector 
by $\vec{a}_\voltens = \tfrac{1}{2}\voltens\ddot\tens{a}$,
which has frame components
$a_\voltens{}^i = \tfrac{1}{2}\volform^{i}{}_{jk} a^{jk}$
with $\volform^{i}{}_{jk} = \delta^{il}\volform_{ljk}$. 
The factor of $\tfrac{1}{2}$ appears so that these two linear operations are inverses of each other 
due to the identity 
$\delta_i^l = \tfrac{1}{2}\volform^l{}_{jk} \volform^{jk}{}_i$. 
Specifically, 
converting a vector $\vec{a}$ to a skew-tensor $\tens{a}_\voltens$ and back to a vector 
$\vec{a}_{\voltens^2} = \tfrac{1}{2}\voltens\ddot\tens{a}_\voltens=\vec{a}$ 
yields the original vector,
and likewise converting a skew-tensor $\tens{a}$ to a vector $\vec{a}_\voltens$ and back to a skew-tensor
$\tens{a}_{\voltens^2}=\voltens\cdot\vec{a}_\voltens=\tens{a}$
yields the original skew-tensor. 
These operations are the three-dimension version of the Hodge dual operator
(see, e.g. \cite{ArnKhe-book,BesFri}). 

The following identities are useful:
\begin{align}
& (\voltens\cdot\vec{a})\ddot(\voltens\cdot\vec{b}) = 2\vec{a}\cdot\vec{b} , 
\label{id:volvec}
\\
& (\voltens\ddot\tens{a})\cdot(\voltens\ddot\tens{b}) = 2\tens{a}\ddot\tens{b} , 
\label{id:voltens}
\end{align}
and
\begin{align}
& \vec{a}\wedge\vec{b}= (\vec{a}\times\vec{b})\cdot\voltens , 
\\
& \vec{a}\times\vec{b}= \tfrac{1}{2}\voltens\ddot(\vec{a}\wedge\vec{b})=(\voltens\cdot\vec{b})\cdot\vec{a} = -(\voltens\cdot\vec{a})\cdot\vec{b} , 
\\
& \vec{a}\wedge\tens{b} = \tfrac{1}{2}(\vec{a}\times\tens{b})\voltens , 
\\
& \vec{a}\times\tens{b} = \vec{a}\cdot(\voltens\ddot\tens{b}) = (\voltens\cdot\vec{a})\ddot\tens{b} , 
\label{id:crossvectens}
\\
& \vec{a}\wedge\vec{b}\wedge\vec{c}= ((\vec{a}\times\vec{b})\cdot\vec{c})\voltens , 
\\
& \vec{a}\wedge(\vec{b}\times\vec{c})= (\vec{a}\cdot\vec{c})\voltens\cdot\vec{b} - (\vec{a}\cdot\vec{b})\voltens\cdot\vec{c} , 
\\
& (\vec{a}\times\vec{b})\cdot\vec{c} = \tfrac{1}{2}(\voltens\ddot(\vec{a}\wedge\vec{b}))\cdot\vec{c} .
\end{align}

\subsection{Differential operations: grad, curl, div, Lie derivative}

The covariant derivative $\nabla$ obeys the Leibniz rule and is a dual version of the familiar gradient operator.
In particular,
$\nabla f$ for any function $f(x)$ is a dual vector field
whose contraction with any vector field $\vec{a}$
yields the directional derivative of $f$ along $\vec{a}$.
It is convenient to convert $\nabla$ into a corresponding contravariant (vectorial) derivative $\vecder$
defined by the property that $\vec{a}\cdot\vecder f$ is the directional derivative of $f$ along $\vec{a}$. 
All differential operations in the tangent space of a Riemannian manifold
can be expressed in terms of this vector operator $\vecder$.

The frame components of $\vecder$ are given by
$\vecder^i = \metr(\e{i},\vecder)=\e{i}\cdot\vecder$, $i=1,2,3$,
where $\e{i}\vecder^i=\vecder$. 
When the manifold is flat, 
these components can be expressed as $\vecder^i = \p_{x^i}$
in Cartesian coordinates $x^i$. 

For a scalar function $a(x)$, 
the basic differential operator is the gradient $\vecder a =\grad a$,
which is a vector function with components $\nabla^i a$. 

For a vector function $\vec{a}(x)$, 
the basic differential operators are the curl $\vecder\times\vec{a}=\curl\vec{a}$
and the divergence $\vecder\cdot\vec{a}=\div\vec{a}$,
which are given by the components 
$(\vecder\times\vec{a})^i = \volform^i{}_{jk}\nabla^j a^k$
with $\volform^{i}{}_{jk} = \delta^{il}\volform_{ljk}$,
and $\vecder\cdot\vec{a} = \delta_{ij}\nabla^i a^j$. 

For a skew-tensor function $\tens{a}(x)$, 
the corresponding differential operators consist of 
$\vecder\times\tens{a}=\curl\tens{a}$
which is a scalar function, given in components by 
$\volform_{ijk}\nabla^i a^{jk}$;  
and $\vecder\cdot\tens{a}=\div\tens{a}$,
which is a vector function, with components 
$(\vecder\cdot\tens{a})^i = \delta_{jk}\nabla^j a^{ki}$. 

The Lie derivative with respect to a vector function $\vec{a}(x)$ is 
defined as follows on scalar functions $b(x)$, vector functions $\vec{b}(x)$, and skew-tensor functions $\tens{b}(x)$:
\begin{align}
& \lieder{\vec{a}} b = \vec{a}\cdot\vecder b , 
\\
& \lieder{\vec{a}} \vec{b} = \vec{a}\cdot\vecder\vec{b} - \vec{b}\cdot\vecder\vec{a} = [\vec{a},\vec{b}]  , 
\\
& \lieder{\vec{a}} \tens{b} = \vec{a}\cdot\vecder\tens{b} - (\tens{b}\cdot\vecder)\vec{a} +((\tens{b}\cdot\vecder)\vec{a})^\t . 
\end{align}
Their frame components are respectively given by 
$a^i\nabla_i b$, 
$a^j\nabla_j b^i - b^j\nabla_j a^i$, 
$a^k\nabla_k b^{ij} - b^{ik}\nabla_k a^j - b^{kj}\nabla_k a^i$. 
An important property is that the Lie derivative of the volume 3-form and volume tensor
are given by 
\begin{equation}\label{Liedervol}
\lieder{\vec{a}}\volform = (\vecder\cdot\vec{a})\volform,
\quad
\lieder{\vec{a}}\voltens = -(\vecder\cdot\vec{a})\voltens . 
\end{equation}
The Lie derivative of the inverse metric tensor $\invmetr$ is given by 
\begin{equation}\label{Liedermetr}
\lieder{\vec{a}}\, \invmetr = -\vecder\odot\vec{a} 
\end{equation}
which is a symmetric derivative. 
In frame components, 
$\lieder{\vec{a}} \delta^{ij} = -(\nabla^i a^j + \nabla^j a^i)$.

\subsection{Lie derivative identities}

From the Lie derivatives \eqref{Liedervol} and \eqref{Liedermetr}, 
the following useful identities are straightforward to prove:
\begin{align}
\lieder{\vec{c}}(\vec{a}\cdot\vec{b}) & 
= \vec{a} \cdot(\lieder{\vec{c}}\vec{b})+ \vec{b} \cdot(\lieder{\vec{c}}\vec{a}) + \symmderu\ddot(\vec{a}\otimes\vec{b}) , 
\\
\lieder{\vec{c}}(\vec{a}\times\vec{b}) & 
= \vec{a} \times(\lieder{\vec{c}}\vec{b}+\symmderu\cdot\vec{b}) -\vec{b} \times(\lieder{\vec{c}}\vec{a}+ \symmderu\cdot\vec{a}) -(\vecder\cdot\vec{c}) \vec{a}\times\vec{b}, 
\\
\lieder{\vec{c}}(\vecder f_1\cdot\vecder f_2) & 
= \vecder f_1\cdot(\lieder{\vec{c}}\vecder f_2)+ \vecder f_2 \cdot(\lieder{\vec{c}}\vecder f_1) 
- \symmderu\ddot(\vecder f_1\otimes\vecder f_2) ,
\\
\lieder{\vec{c}}(\vecder f_1\times\vecder f_2) & 
=  \vecder f_1 \times\vecder(\lieder{\vec{c}} f_2) -\vecder f_2 \times\vecder(\lieder{\vec{c}} f_1) -(\vecder\cdot\vec{c}) \vecder f_1\times\vecder f_2 . 
\end{align}

\section{Transcription between vector/tensor calculus\\ and differential forms}\label{dictionary}

In any three-dimensional Riemannian manifold, 
differential forms are a counterpart of vectors $\vec{a}$, skew-tensors $\tens{a}$, and totally-antisymmetric tensors $a\voltens$.
Specifically, 
a 1-form is a linear map from vectors into scalars; 
a 2-form is a linear map from skew-tensors into scalars;
and a 3-form is a linear map from totally-antisymmetric tensors into scalars.

With respect to a frame given by three orthonormal vectors $\e{i}$, $i=1,2,3$, 
the components of a 1-form $\alpha$ are given by 
$\alpha_i=\alpha(\e{i})$,
and similarly the components of a 2-form $\beta$ are given by 
$\beta_{ij}=-\beta_{ji}=\beta(\e{i}\wedge\e{j})$. 
The components of a 3-form $\gamma$ are a scalar multiple $\tilde\gamma$ of the components of the volume form $\volform$, 
namely 
$\gamma_{ijk} = \tilde\gamma\volform_{ijk} = \gamma(\e{i}\wedge\e{j}\wedge\e{k})$;
in particular, $\gamma = \tilde\gamma\volform$.

There is a one-to-one correspondence 
between 1-forms $\alpha$ and vectors $\vec{a}$, 
which arises from the metric.
In frame components: 
\begin{equation}\label{vec=1form}
\alpha_i = \delta_{ij}a^i,
\quad
a^i = \delta^{ij}\alpha_j . 
\end{equation}
This correspondence extends to 2-forms $\beta$ and skew-tensors $\tens{a}$:
\begin{equation}\label{tens=2form}
\beta_{ij} = \delta_{ik}\delta_{jl} a^{kl},
\quad
a^{ij} = \delta^{ik}\delta^{jl}\beta_{kl} . 
\end{equation}

Another one-to-one correspondence arises from the volume tensor $\voltens$ and the volume form $\volform$. 
Specifically, 
a 1-form $\alpha$ can be converted into a skew-tensor $\tens{a}$ 
given by 
\begin{subequations}\label{tens=1form}
\begin{equation}\label{tens1form}
a^{ij} = \volform^{ijk}\alpha_k . 
\end{equation}
Correspondingly, a skew-tensor $\tens{a}$ can be converted into a 1-form $\alpha$ 
given by 
\begin{equation}\label{1formtens}
\alpha_i = \tfrac{1}{2}\volform_{ijk} a^{jk} . 
\end{equation}
\end{subequations}
The factor of $\tfrac{1}{2}$ appears here so that these two linear operations are inverses of each other through the identity 
$\delta_k^l = \tfrac{1}{2}\volform_{ijk}\volform^{ijl}$. 
In a similar way,
through the previous correspondence between skew-tensors and 2-forms, 
a 2-form $\beta$ can be converted into a vector $\vec{a}$ 
given by 
\begin{subequations}\label{vec=2form}
\begin{equation}\label{vec2form}
a^i = \tfrac{1}{2}\volform^{ijk}\beta_{jk} , 
\end{equation}
and conversely 
\begin{equation}\label{2formtens}
\beta_{ij} = \volform_{ijk} a^k
\end{equation}
\end{subequations}
converts a vector $\vec{a}$ back into a 2-form $\beta$. 

It is natural to extend these correspondences by 
first defining a 0-form as a scalar $a$,
and then using the correspondence between scalars and totally-antisymmetric tensors to convert a 0-form $a$ into a totally-antisymmetric tensor $a\voltens$
having components $a^{ijk} = a\volform^{ijk}$. 
Conversely, a totally-antisymmetric tensor is converted back into a 0-form 
given by $a=\tfrac{1}{6}\volform_{ijk} a^{ijk}$. 

Next we recall the basic wedge product and exterior derivative operations on differential forms, 
along with contraction between differential forms and vectors and tensors. 
We show how these operations mirror the basic algebraic and differential operations on vectors and tensors. 

\subsection{Operations on differential forms}

The wedge product of 1-forms corresponds to the exterior product of vectors,
and the wedge product of 1-form and 2-form corresponds to the exterior product of a vector and a skew-tensor. 
In particular, given two 1-forms $\alpha$ and $\mu$, 
their wedge product $\alpha\wedge\mu=\beta$ is a 2-form 
with frame components 
$\beta_{ij}= \alpha_i\mu_j- \mu_i\alpha_j$. 
Likewise, given a 1-form $\alpha$ and a 2-form $\beta$, 
their wedge product $\alpha\wedge\beta=\beta\wedge\alpha=\gamma$ is a 3-form 
with frame components 
$\gamma_{ij}= \alpha_i\beta_{jk} + \alpha_j\beta_{ki} +\alpha_k\beta_{ij}$. 

The wedge product of two 2-forms vanishes because there cannot exist a totally-antisymmetric tensor of rank 4 in three dimensions. 

Contraction of a vector $\vec{a}$ with a 1-form $\alpha$ is defined by 
$\vec{a}\hook\alpha = a^i\alpha_i$ in frame components. 
Similarly, contraction of a vector $\vec{a}$ with, respectively, a 2-form $\beta$ and a 3-form $\gamma=a\volform$ is defined by 
$\vec{a}\hook\beta = a^i\beta_{ij}$ 
and $\vec{a}\hook\gamma = a^i\gamma_{ijk}$. 
A skew-tensor $\tens{a}$ can be contracted with, respectively, a 2-form $\beta$ and a 3-form $\gamma$ by 
$\tens{a}\hook\beta = a^{ij}\beta_{ij}$ 
and 
$\tens{a}\hook\gamma = a^{ij}\gamma_{ijk}$.

The basic derivative operator on differential forms is the exterior derivative $\d$
which corresponds to a curl on vectors and skew-tensors,
and a gradient on scalars. 
In frame components, $\d$ is defined on 
0-forms $a$, 1-forms $\alpha$, 2-forms $\beta$ 
by:
\begin{align}
& (\d a)_i=\nabla_i a , 
\label{exact1form}\\
& (\d\alpha)_{ij}=\nabla_i \alpha_j - \nabla_j \alpha_i , 
\label{exact2form}\\
& (\d\beta)_{ijk}= \nabla_i \beta_{jk}+ \nabla_j \beta_{ki} +\nabla_k \beta_{ij} , 
\label{exact3form}
\end{align}
where $\nabla_i$ denotes the frame components of the covariant derivative. 
These differential forms are called exact. 
The exterior derivative of any exact differential form vanishes, 
due to the property $\d^2=0$, corresponding to the vector calculus properties $\div\curl=0$ and $\curl\grad=0$. 

Through the two correspondences \eqref{vec=1form} and \eqref{tens=1form}
between 1-forms and vectors, 
an exact 1-form \eqref{exact1form} represents a gradient vector $\nabla^i a$ 
and a divergence of a totally-antisymmetric tensor 
$\nabla_k(\volform^{ijk}a)=\volform^{ijk}\nabla_k a$. 
Likewise, 
an exact 2-form \eqref{exact2form} represents an antisymmetric derivative of a vector $\nabla^i a^j - \nabla^j a^i$ 
and a curl of a vector $2\volform^i{}_{jk}\nabla^j a^k$,
by the correspondences \eqref{tens=2form} and \eqref{vec=2form}. 
This curl can also be expressed as a divergence of a skew-tensor 
$\nabla_j(\volform^{ij}{}_k a^k)$. 
In a similar way, 
an exact 3-form \eqref{exact3form} represents 
an antisymmetric derivative of a skew-tensor 
$\nabla^i a^{jk}+ \nabla^j a^{ki} +\nabla^k a^{ij}$
and a divergence of a vector 
$3\nabla_i(\volform^i{}_{jk} a^{jk})$. 

Another derivative operator on differential forms is the Lie derivative
with respect to any vector field $\vec{a}$:
\begin{align}
\lieder{\vec{a}} a & = \vec{a}\hook(\d a) , 
\\
\lieder{\vec{a}}\alpha & = \vec{a}\hook(\d\alpha) + \d(\vec{a}\hook\alpha) , 
\\
\lieder{\vec{a}}\beta & = \vec{a}\hook(\d\beta) + \d(\vec{a}\hook\beta) , 
\\
\lieder{\vec{a}}\gamma & = \d(\vec{a}\hook\gamma) .
\end{align}
Here, $\hook$ denotes the contraction (interior product) between a vector and a differential form. 
The Lie derivative has the property that it commutes with the  
exterior derivative $\d$:
\begin{equation}
[\lieder{\vec{a}},\d]=0 . 
\end{equation}

For a flat Riemannian manifold,
when the exterior derivative is applied to Cartesian coordinates $x^i$, $i=1,2,3$, 
this yields a basis of three 1-forms $(\d x^1,\d x^2,\d x^3)$, 
denoted $\d\x$. 
The wedge product of the these 1-forms then yields a basis of three 2-forms:
$(\d x^1\wedge\d x^2,\d x^2\wedge\d x^3,\d x^3\wedge\d x^1)$, 
denoted $\d\x\wedge\d\x$. 
Every 1-form $\alpha$ and 2-form $\beta$ can be expanded in this basis 
in terms of Cartesian components
\begin{equation}
\alpha=\alpha_i\d x^i, 
\quad
\beta=\beta_{ij}\d x^i\wedge\d x^j .
\end{equation}

\subsection{Euler's equations using differential forms}

Euler's equations \eqref{eulereqn} for inviscid fluid flow 
can be converted into differential forms
by using the frame components of the fluid velocity
$\u = u^i\e{i}$
where $\e{i}$, $i=1,2,3$, is any frame consisting of orthonormal vectors in $\Rnum^3$.

Using Cartesian coordinates $x^i$,
write 
\begin{equation}\label{vel=1form}
\mbs u=\delta_{ij}u^i\d x^j=\u\cdot\d\x
\end{equation} 
for the 1-form corresponding to the fluid velocity. 
The curl of the fluid velocity corresponds to the exterior derivative of $u$:
\begin{equation}\label{vort=2form}
\mbs\omega = \d\mbs u = \volform_{ijk}(\vecder\times\u)^k \d x^i\wedge\d x^j
= \volform_{ijk}\omega^k \d x^i\wedge\d x^j 
\end{equation} 
where the vorticity vector $\vort =\vecder\times\u$ is identified with 
the vorticity 2-form $\mbs\omega$ through the correspondence \eqref{vec=2form}.
The divergence free property of $\vort$ then corresponds to $\d\mbs\omega=0$. 

To express Euler's equations \eqref{eulereqn} in terms of $\mbs u$ and $\mbs\omega$, 
we convert the vector calculus identity \eqref{skewderuident} into a corresponding 1-form identity
\begin{equation}
\u\cdot\vecder u = \tfrac{1}{2}\d(|\u|^2) +\u\hook\mbs\omega .
\end{equation}
This yields \cite{App,Anc13,WebDasMcKHuZan,BesFri}
\begin{equation}\label{vel1formeqn}
\mbs u_t = -\u\hook\mbs\omega -\tfrac{1}{2}\d(|\u|^2) -(1/\rho)\d p .
\end{equation} 
Hence we have
\begin{equation}\label{vort2formeqn}
\mbs\omega_t = -\d(\u\hook\mbs\omega) +(1/\rho^2)\d\rho\wedge\d p ,
\end{equation} 
which is the 2-form version of the vorticity equation \eqref{vorteqncurlform}. 
Likewise,
the entropy equation \eqref{entropyeqn} becomes
\begin{equation}\label{entr0formeqn}
S_t + \u\hook\d S =0 .
\end{equation} 

These equations \eqref{vel1formeqn}--\eqref{entr0formeqn}
have an elegant formulation using the advective Lie derivative \eqref{advLieder}:
\begin{align}
\Dt\mbs u & = \d(\tfrac{1}{2}|\u|^2 -E -p/\rho) +T\d S , 
\\ 
\Dt\mbs\omega & = \d T\wedge \d S , 
\quad
\mbs\omega =\d\mbs u , 
\\
\Dt S & =0 ,
\end{align}
where $T$ is the fluid temperature and $E$ is the internal energy of the fluid,
given by the thermodynamic relation \eqref{thermorel}. 
This fluid velocity equation for $\mbs u$ 
is the differential-form version of Weber's transformation 
\cite{Web,App,WebDasMcKHuZan,BesFri}.

\subsection{Invariants}

Invariant vectors $\vec{\inv}$ and invariant skew-tensors $\tens{\inv}$ 
respectively correspond to invariant 2-forms $\kappa=\tfrac{1}{2}\rho\vec{\inv}\hook\volform$ 
and invariant 1-forms $\lambda=\tfrac{1}{2}\rho\tens{\inv}\hook\volform$,
\begin{equation}
\Dt\kappa=\p_t\kappa + \Lu\kappa =0,
\quad
\Dt\lambda=\p_t\lambda + \Lu\lambda =0 .
\end{equation}
The converse is $\vec{\inv} = (1/\rho)\voltens\hook\kappa$ 
and $\tens{\inv}= (1/\rho)\voltens\hook\lambda$. 
This correspondence arises from the property that 
the densitized volume form $\rho\volform$ 
and the densitized volume tensor $(1/\rho)\voltens$ are invariants, 
$\Dt(\rho\volform)=0$ and $\Dt((1/\rho)\voltens)=0$,
whereas the metric $\metr$ is not an invariant as show by its advection property \eqref{metrtransport}. 

In particular,
consider a Cartesian frame $\e{i}$ which is dual to $\d x^i$:
$\e{i}\hook\d x^j =\delta_i^j$. 
Then 
$\vec{\inv} = \inv^i\e{i}$ is an invariant vector iff 
$\kappa= \kappa_{ij}\d x^j\wedge\d x^k$ is an invariant 2-form, 
where $\kappa_{ij} = \tfrac{1}{2}\rho\volform_{ijk}\inv^k$
and $\inv^i=(1/\rho)\volform^{ijk}\kappa_{jk}$. 
The component form of the invariance property is given by 
$(\Dt\vec{\inv})^i = \p_t\inv^i + u^j\nabla_j\inv^i - \inv^j\nabla_j u^i=0$
and
$(\Dt\kappa)_{ij} = \p_t\kappa_{ij} + u^k\nabla_k\kappa_{ij} + \kappa_{ki}\nabla_j u^k- \kappa_{kj}\nabla_i u^k =0$.
Similarly, 
$\tens{\inv} = \inv^{ij}\e{i}\otimes\e{j}$ is an invariant skew-tensor iff 
$\lambda= \lambda_i\d x^i$ is an invariant 1-form, 
where $\lambda_i = \tfrac{1}{2}\rho\volform_{ijk}\inv^{jk}$
and $\inv^{ij}= (1/\rho)\volform^{ijk}\lambda_k$. 
The component form of the invariance property is given by 
$(\Dt\tens{\inv})^{ij} = \p_t\inv^{ij} + u^k\nabla_k\inv^{ij} - \inv^{kj}\nabla_k u^i + \inv^{ki}\nabla_k u^j=0$
and
$(\Dt\lambda)_i = \p_t\lambda_i + u^k\nabla_k\lambda_i + \lambda_k\nabla_i u^k=0$. 

In the terminology and notation of \Ref{TurYan}, 
scalar and vector invariants 
are respectively called a Lagrangian-type invariant $I=\inv$ 
and a field-type invariant $\vec{J}=\vec{\inv}$, 
while skew-tensor invariants $\tens{\inv}$ appear only in the form of a vector
$\vec{S}= \rho\voltens\ddot\tens{\inv}$. 
Such vectors are a kind of dual of skew-tensor invariants
but are not themselves invariant since 
$\Dt\vec S= -\vec S\cdot\symmderu \neq 0$
follows from $S^i =  \rho\delta^{il}\volform_{ljk}\inv^{jk}$ 
by using the transport properties \eqref{metrtransport} and \eqref{voltransport} of $\delta^{ij}$ and $\volform_{ijk}$. 
The proper way to view $\vec S$ is that its associated 1-form 
$S_i = \rho\volform_{ijk} \inv^{jk}$ is an actual invariant.

\subsection{A generating set of material operators for advected differential forms}

The material operators shown in Theorem~\ref{materialops} have a simple formulation in terms of differential forms. 

\emph{Material differential operators}:

If $\mu$ is an advected 0-form, 1-form, or 2-form,
then $\d\mu$ is an advected 1-form, 2-form, or 3-form. 

If $\mu$ is an advected 0-form, 1-form, 2-form, or 3-form, 
and $\vec{\inv}$ is an invariant vector, 
then so is $\lieder{\vec{\inv}}\mu = \vec{\inv}\hook\d\mu + \d(\vec{\inv}\hook\mu)$. 

\emph{Material algebraic operators}:

If $\mu$ is an advected 1-form, 2-form, or 3-form,
and $\vec{\inv}$ is an invariant vector, 
then  $\vec{\inv}\hook\mu$ is an advected 0-form, 1-form, or 2-form.

If $\mu$ and $\nu$ are advected 1-forms, 
then $\mu\wedge\nu$ is an advected 2-form. 
If $\mu$ is an advected 1-form and $\nu$ is an advected 2-form, 
then $\mu\wedge\nu$ is an advected 3-form. 

If $\mu$ is an advected 2-form, 
then $(1/\rho)\voltens\hook\mu$ is an invariant vector. 
Conversely, if $\vec{\inv}$ is an invariant vector, 
then $\rho\volform\hook\vec{\inv}$ is an advected 2-form. 

If $\mu$ is an advected 3-form, 
then $(1/\rho)\voltens\hook\mu$ is an advected 0-form, namely, an invariant scalar. 
Conversely, if $\inv$ is an invariant scalar, 
then $\rho\inv\volform$ is an advected 3-form.

\end{document}